\documentclass[a4paper, twocolumn=off, twoside=on]{scrartcl} 

\KOMAoptions{fontsize=11pt} %
\KOMAoptions{BCOR=0mm}		%

\usepackage{amsmath}
\usepackage{amsthm}
\usepackage[fixamsmath]{mathtools}
\mathtoolsset{showonlyrefs=true}
\usepackage[english,ngerman]{babel}
\usepackage[long,nodayofweek]{datetime}
\usepackage{enumitem}

\usepackage[usenames,dvipsnames]{xcolor}
\usepackage{siunitx} 
\usepackage{url}
\usepackage{float}
\usepackage{caption}
\usepackage{subcaption}
\usepackage{array}
\usepackage{booktabs}
\usepackage{authblk}
\usepackage{xfrac}
\usepackage{inputenc}
\usepackage[T1]{fontenc}
\usepackage[final,babel=true,expansion=true,protrusion=true,spacing=true,kerning=true]{microtype}

\usepackage[adobe-utopia]{mathdesign} 
\usepackage[ttscale=.875]{libertine}    

\usepackage[hidelinks]{hyperref}
\usepackage[a4paper,inner=3.0cm,outer=3cm,top=2cm,bottom=1.5cm, includeheadfoot]{geometry}

\usepackage{tikz}
\usepackage{pgfplots}
\pgfplotsset{compat=newest}
\usetikzlibrary{positioning,calc}
\usetikzlibrary{fit}
\usepgfplotslibrary{fillbetween}
\usepgfplotslibrary{groupplots}

\usepgfplotslibrary{external}
\tikzsetexternalprefix{figures/}

\newcommand{\bR}{\mathbb{R}}
\newcommand{\bN}{\mathbb{N}}

\newcommand{\cA}{\mathcal{A}}

\newcommand{\cL}{\mathcal{L}}
\newcommand{\cP}{\mathcal{P}}
\newcommand{\cQ}{\mathcal{Q}}
\newcommand{\cR}{\mathcal{R}}

\newcommand{\cU}{\mathcal{U}}

\newcommand{\cX}{\mathcal{X}}
\newcommand{\cY}{\mathcal{Y}}
\newcommand{\cZ}{\mathcal{Z}}

\newcommand{\phasel}{-}
\newcommand{\phaser}{+}
\newcommand{\phaselr}{\mp}
\newcommand{\spx}{x} %
\newcommand{\vu}{\vec{u}} %
\newcommand{\ix}{\vec{x}} %
\newcommand{\oy}{\vec{y}} %
\newcommand{\cres}{CRes} %
\newcommand{\cadl}{CAL} %

\newcommand{\vW}{\vec{w}}
\newcommand{\vb}{\vec{b}}
\newcommand{\vZ}{\vec{z}}

\renewcommand{\vec}[1]{\boldsymbol{#1}}

\DeclarePairedDelimiterX{\jump}[1]{[}{]}{\hspace{-0.2em}\delimsize[ {#1} \delimsize]\hspace{-0.2em}}
\DeclarePairedDelimiterX{\mean}[1]{\{}{\}}{\hspace{-0.25em}\delimsize\{ {#1} \delimsize\}\hspace{-0.25em}}

\DeclareMathOperator{\ddd}{d\!}

\DeclarePairedDelimiter\abs{\lvert}{\rvert}

\DeclarePairedDelimiterX{\norm}[1]{\lVert}{\rVert}{
\ifblank{#1}{\:\cdot\:}{#1}
}

\usepackage[autostyle]{csquotes}
\usepackage[backend=biber, style=numeric, sorting=nyt, eprint=true, doi=true, isbn=false, url=false, maxcitenames=2, maxbibnames=5, giveninits=true]{biblatex}

\DeclareFieldFormat{title}{\mkbibemph{#1}} 
\DeclareFieldFormat[article]{title}{\mkbibemph{#1\isdot}} 
\DeclareFieldFormat[inbook]{title}{\mkbibemph{#1\isdot}} 
\DeclareFieldFormat[incollection]{title}{\mkbibemph{#1\isdot}} 
\DeclareFieldFormat[inproceedings]{title}{\mkbibemph{#1\isdot}} 
\DeclareFieldFormat[patent]{title}{\mkbibemph{#1\isdot}} 
\DeclareFieldFormat[thesis]{title}{\mkbibemph{#1\isdot}} 
\DeclareFieldFormat[unpublished]{title}{\mkbibquote{#1\isdot}} 
\DeclareFieldFormat[suppbook]{title}{#1} 
\DeclareFieldFormat[suppcollection]{title}{#1} 
\DeclareFieldFormat[suppperiodical]{title}{#1}
\DeclareFieldFormat{url}{url: \url{#1}}

\AtEveryBibitem{%
\clearfield{url}%
  \clearlist{language}%
  \clearfield{type}%
  \clearfield{pages}%
  \clearfield{month}%
  \clearfield{day}%
  \clearfield{issue}%
  \clearfield{pagetotal}%
}

\AtEveryCitekey{%
\clearfield{url}%
\clearfield{pages}%
\clearfield{month}%
\clearfield{day}%
\clearfield{issue}%
\clearfield{pagetotal}%
\clearfield{note}%
\clearlist{language}%
\clearfield{type}%
\clearname{translator}%
\clearlist{publisher}%
\clearlist{location}%
}

\renewbibmacro*{doi+eprint+url}{%
    \printfield{doi}%
    \iffieldundef{doi}{%
    \newunit\newblock%
    \iftoggle{bbx:eprint}{%
        \usebibmacro{eprint}%
    }{}%
    \newunit\newblock%
    \iffieldundef{eprint}{%
        \usebibmacro{url+urldate}}%
        {}%
    }{}    
    }

\theoremstyle{remark}
\newtheorem*{remark}{Remark}

\KOMAoptions{parskip=half}
\addtokomafont{section}{\centering}
\addtokomafont{titlehead}{}
\addtokomafont{subject}{\normalfont\sffamily\normalsize}
\addtokomafont{title}{\normalfont\sffamily\normalfont\sffamily\Large}
\addtokomafont{subtitle}{\normalfont\sffamily\large}
\addtokomafont{author}{\normalfont\sffamily\large}
\addtokomafont{date}{\normalfont\sffamily\large}
\addtokomafont{publishers}{\normalfont\sffamily\normalsize}

\setenumerate[1]{label={\arabic*}.,  leftmargin=*}
\setitemize{leftmargin=*}
\setlist{rightmargin=1em}

\allowdisplaybreaks[1]  %

\usepackage[clines, plainfootsepline, automark]{scrlayer-scrpage}

\clearscrheadfoot

\ihead[]{}
\ohead[]{\pagemark}
\ifoot[]{}
\ofoot[]{}

\setkomafont{pagehead}{%
\normalfont\footnotesize\itshape}
\setkomafont{pagefoot}{%
\normalfont\footnotesize}
\setkomafont{pagenumber}{%
\normalfont\normalsize\itshape}

\makeatletter
\g@addto@macro\bfseries{\boldmath}
\makeatother

\newcommand*\circled[1]{\tikzset{external/export next=false}\tikz[baseline=(char.base)]{
            \node[shape=circle,draw,inner sep=1pt] (char) {\small{}#1};}}
  
\addto\extrasenglish{%

}

\begin{document} 

\selectlanguage{english}

\title{Constraint-Aware Neural Networks for Riemann Problems}  
\renewcommand\Affilfont{\itshape\small}
\renewcommand{\Authands}{ and }

\author[1]{Jim Magiera}
\author[2]{Deep Ray}
\author[2]{Jan S. Hesthaven}
\author[1]{Christian Rohde}
\affil[1]{University of Stuttgart, Institute of Applied Analysis and Numerical Simulation}
\affil[2]{\'Ecole Polytechnique F\'ed\'erale de Lausanne (EPFL), Chair of Computational Mathematics and Simulation Science}
\date{\small{}April 2019}

\maketitle

\KOMAoptions{abstract=true}
\begin{abstract}
Neural networks are increasingly used in complex (data-driven) simulations as surrogates or for accelerating the computation of classical surrogates.  
In many applications physical constraints, such as mass or energy conservation, must be satisfied to obtain reliable results. 
However, standard machine learning algorithms are generally not tailored to respect such constraints.
\newline
We propose two different strategies to generate constraint-aware neural networks.
We test their performance in the context of front-capturing schemes for strongly nonlinear wave motion in compressible fluid flow.
Precisely, in this context so-called Riemann problems have to be solved as surrogates.
Their solution describes the local dynamics of the captured wave front in numerical simulations.
Three model problems are considered: a cubic flux model problem, an isothermal two-phase flow model, and the Euler equations.
We demonstrate that a decrease in the constraint deviation correlates with low discretization errors for all model problems, in addition to the structural advantage of fulfilling the constraint.
\end{abstract}

\section{Introduction}

Machine-learned (surrogate) models have gained great popularity in various fields of research and application, particularly in the context of dynamical systems \cite{brunton:sparse:2016,mishra:mlda:2018,brunton_kutz_2019} and  partial differential equations \cite{golak:annpde:2007,rudd:cint:2015,tompson:cnn:2017,ray.hesthaven:artificial:2018,zichao2018,pang:nngp:2019}. Most physical systems are subject to secondary or inferred constraints, such as conservation of mass and energy for inviscid flows, satisfaction of maximum principles or the Rankine--Hugoniot conditions in the context of hyperbolic partial differential equations. Thus, it becomes crucial that the surrogate models satisfy such constraints to faithfully represent the physical behavior of the underlying system.

Some advances have been made to develop constraint-aware methods using traditional machine learning algorithms, such as positivity preservation of output variables \cite{derossi.perracchione:positive:2017}, or generating divergence-free vector fields \cite{narcowich.ward.ea:divergence:2007}. 
With the growing capabilities of deep learning \cite{lecun.bengio.ea:deep:2015} it is highly interesting to develop methods that are applicable for neural networks.
Some constraints, like bounded output variables are trivially accomplished by choosing an appropriate activation function in the output layer (refer to \autoref{sec:neural_networks}).
On the other hand, more complex constraints are still an open challenge.
In recent years there have been some efforts to resolve this problem.
In \cite{lee.wick.ea:gradient:2017} unconstrained neural networks are combined with an optimization algorithm during test time to resolve constraints in the context of parsing problems.
A probabilistic approach to affine-linear constraints was proposed in \cite{pathak.kraehenbuehl.ea:constrained:2015}.
In \cite{marquez-neila.salzmann.ea:imposing:2017} the performance of neural networks that resolve constraints exactly was compared to networks that satisfy them only approximately.

We are convinced that the design of constraint-aware machine learning tools depends crucially on the underlying mathematical  model and the chosen discretization method.
In this contribution we focus on the seminal wave-tracking problem in compressible fluid flow governed by hyperbolic conservation laws.
More specifically, we focus on learning Riemann solvers for hydrodynamical interfaces, while taking into account the conservation of mass and momentum.
Computing Riemann solutions is the key operation that has to be performed many times in front-capturing schemes.
\newline
Up to our knowledge nothing has been done on constraint-aware learning methods in the field of conservation laws and their discretization techniques.

The rest of the paper is structured as follows:
In \autoref{sec:problem_description}, we present the problem description and a general system of conservation laws which will be our primary model problem. 
A brief summary of neural networks is given in \autoref{sec:neural_networks}, with a specific focus on multilayer perceptrons.
In \autoref{sec:constraint-aware-methods} two constraint-aware learning methods are presented.
The first method (\cres) is an analytic approach based on the form of the constraint and the underlying equations.
The second method (\cadl) is more general, as the constraint is added as a penalty term during the training process.
Three case study model problems are described in \autoref{sec:models}, forming the foundation of the numerical performance tests carried out in \autoref{sec:results}.
The technical details of the algorithms involved in the tests, such as the front-capturing finite volume scheme, are given in the preceding \autoref{sec:technical_details}.
Finally, \autoref{sec:summary} is dedicated to the concluding discussion.

\section{Problem Description}
\label{sec:problem_description}

We consider the Cauchy problem for a generic hyperbolic system of conservation laws in one space dimension
\begin{align} \label{eq:conservation_law}
\begin{aligned}
 \partial_t \vu + \partial_\spx f(\vu) &= \vec{0}  && \text{ in } \bR \times (0,t_{\mathrm{end}}), \\
 \vu(\:\cdot\:,0) &= \vu_0 && \text{ in } \bR,
\end{aligned}
\end{align}
where $\vu \colon \bR \times [0,t_{\mathrm{end}}) \to \cU$ is unknown, with the open set $\cU \subset \bR^m$ denoting the state space of the system.
By $f \in C^1(\cU, \bR^m)$ we denote the given flux function.
It is well known that for nonlinear flux functions $f$ classical solutions may break down after a finite time, even for smooth initial datum $\vu_0 \colon \bR \to \cU$.
Therefore the principle of weak solvability is applied to allow for discontinuous solutions.
\newline
A function $\vu \in L^{\infty}(\bR \times [0,t_{\mathrm{end}}),  \cU)$ is called a weak solution of the Cauchy problem \eqref{eq:conservation_law} with initial data $\vu_0 \in L^{\infty}(\bR, \cU)$ if
\begin{align} \label{eq:weak_solution}
 \int^{t_\mathrm{end}}_{0} \int_{\bR} \bigl(\vu \partial_t \psi + f(\vu) \partial_\spx \psi \bigr)~\ddd \spx \ddd t = - \int_{\bR} \psi(\spx,0) \vu_0(\spx)~\ddd \spx
\end{align}
holds for all compactly supported test functions $\psi \in C^{\infty}_{0}(\bR \times [0,t_{\mathrm{end}}))$.

However, weak solutions \eqref{eq:weak_solution} are not unique,
and require the prescription of additional criteria to single out a physically relevant solution.
Alternatively, a unique weak solution can be selected as the limit of solutions of   versions of \eqref{eq:weak_solution} that account for microscopic effects.
Classical examples include dissipative (viscous, diffuse--dispersive, relaxation) approximations, kinetic approaches via Boltzmann hierarchies or approximation by  averaged trajectories of molecular dynamics.

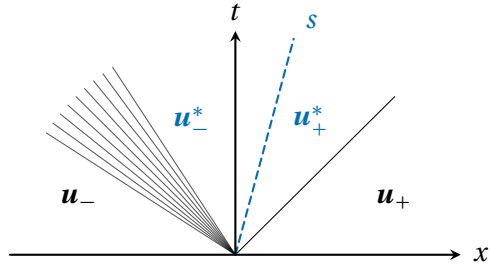
\begin{figure}[ht]
 \centering
\resizebox{0.45\columnwidth}{!}{
\tikzsetnextfilename{wave_pattern_euler}
\begin{tikzpicture}[scale=3.0, font=\sffamily]

\tikzset{coordinateaxis/.style={->, >=stealth, thick, line join=round, line cap=round}}
\tikzset{shockline/.style={-, line join=round, line cap=round, opacity=1.0}}
\tikzset{rarefactionline/.style={-, line join=round, line cap=round, opacity=0.66}}

\draw[shockline] (0,0) -- ({cos(45)}, {sin(45)});
\draw[shockline, NavyBlue, thick, densely dashed] (0,0) -- ({cos(75)}, {sin(75)}) node[pos=1.0, above right]{$s$};

\def \n {9}
\foreach \s in {1,...,\n}
{
	\draw[rarefactionline] (0,0) -- ({cos(120 + 3 * \s)}, {sin(120 + 3 * \s)});
}

\draw[coordinateaxis] (-1,0) -- (1,0) node[pos=1.0, right] {$\spx$};
\draw[coordinateaxis] (0,0) -- (0,1) node[pos=1.0, above] {$t$};

\node at (-0.7, 0.25) {$\vu_{\phasel}$};
\node at (0.7, 0.25) {$\vu_{\phaser}$};
\node[NavyBlue] at (-0.20, 0.6) {$\vu^{*}_{\phasel}$};
\node[NavyBlue] at (0.33, 0.6) {$\vu^{*}_{\phaser}$};

\end{tikzpicture}
}
\caption{Example of a Riemann solution consisting of (from left to right) a rarefaction wave, a contact discontinuity and a shock wave. The wave of interest may be in this case the contact discontinuity.}
\label{fig:wave_pattern:euler}
\end{figure}

In the following, we assume that for $\vu_{\phasel} \in \cP_{\phasel}$ and $\vu_{\phaser} \in \cP_{\phaser}$ chosen from some suitable subsets $\cP_{\phasel}, \cP_{\phaser} \subset \cU$, the unique weak solution to the Riemann initial data
\begin{align} \label{eq:riemannproblem}
 \vu_0(\spx) = \begin{cases}
           \vu_{\phasel} & \text{ for } \spx < 0, \\
           \vu_{\phaser} & \text{ for } \spx > 0,
          \end{cases}
\end{align}
exists.
The solution of \eqref{eq:riemannproblem} typically consists of a composition of elementary waves, such as shocks, contacts and rarefactions.
Out of them, we are interested in the dynamics of one specific, discontinuous \emph{wave of interest}, which may represent a phase boundary, a fluid interface, etc.
We refer to \autoref{fig:wave_pattern:euler} for an illustration of a typical wave pattern including a wave of interest.
The wave of interest is represented by its two adjacent states $\vu^{*}_{\phasel} \in \cP_\phasel$, $\vu^{*}_{\phaser} \in \cP_\phaser$ and the wave speed $s \in \bR$.
The tuple $(\vu^{*}_{\phasel}, \vu^{*}_{\phaser}, s)$ corresponds to a traveling wave
\begin{align} \label{eq:woi}
\vu^*(\spx,t) =
\begin{cases}
           \vu^{*}_{\phasel} & \text{ for } \spx < st, \\
           \vu^{*}_{\phaser} & \text{ for } \spx > st,
          \end{cases}
\end{align}
which is itself a weak solution of \eqref{eq:conservation_law} with Riemann datum consisting of the two adjacent states.
\newline
The wave of interest plays a significant role in the dynamics of  \eqref{eq:conservation_law}.
Consequently, it is important to resolve it accurately in numerical simulations.
This can be achieved by using front-tracking algorithms such as the ghost-fluid method \cite{fedkiw.aslam.ea:non:1999,fechter.munz.ea:approximate:2018},
or front-capturing algorithms such as moving mesh methods \cite{chalons.rohde.ea:finite:2017}.
For each of these numerical algorithms it is essential to describe the dynamics of the wave of interest precisely.
In this work we will focus on a moving mesh algorithm, although the same set of issues is relevant to alternative methods.

The moving-mesh algorithm approximates the location of the wave of interest as an interface between two (moving) cells.
This approximation requires the wave speed $s$ and the trace states $\vu^{*}_{\phasel}$, $\vu^{*}_{\phaser}$ at the wave front, which are usually obtained via a Riemann solver for \eqref{eq:conservation_law} with the Riemann initial data \eqref{eq:riemannproblem} (see  \autoref{fig:wave_pattern:euler}).
Such Riemann solvers can be seen as mappings of the type
\begin{align} \label{eq:riemannsolver:generic}
 \cR \colon \cP_{\phasel} \times \cP_{\phaser} \to  \bR^m \times \bR^m \times \bR:
 (\vu_{\phasel}, \vu_{\phaser}) \mapsto (\vu^{*}_{\phasel}, \vu^{*}_{\phaser}, s).
\end{align}
The tuples $(\vu^{*}_{\phasel}, \vu^{*}_{\phaser}, s)$ returned by \eqref{eq:riemannsolver:generic} define a traveling wave \eqref{eq:woi} which fulfills the Rankine--Hugoniot condition
\begin{align} \label{eq:rh_condition}
 s \, (\vu^{*}_{\phasel} - \vu^{*}_{\phaser}) = f(\vu^{*}_{\phasel}) - f(\vu^{*}_{\phaser}).
\end{align}
However, \eqref{eq:rh_condition} might not hold if the solution of the Riemann problem \eqref{eq:riemannproblem} is computed from dissipative approximations, such as the Navier--Stokes equations or particle simulations \cite{magiera.rohde:particle:2018}.
Especially in the former case, the solution might be corrupted by noise, due to fluctuations in the particle distribution, and therefore \eqref{eq:rh_condition} might not hold true.
\newline
Furthermore, the computational cost associated with using accurate Riemann solvers can be quite high, particularly when applied in a multidimensional numerical tracking scheme, where $\cR$ needs to be evaluated across several mesh interfaces describing the wave of interest, and at each time step.

To address the high computational costs, it is attractive to build reduced surrogate models of the Riemann solver \eqref{eq:riemannsolver:generic} by employing machine learning algorithms such as support vector machines \cite{kissling.rohde:computation:2015} or artificial neural networks.

\begin{figure}[ht]
 \centering
\resizebox{0.8\columnwidth}{!}{
\tikzsetnextfilename{scheme_overview}
\pgfdeclarelayer{bg}%
\pgfdeclarelayer{mybg}%
\pgfdeclarelayer{fg}%
\pgfsetlayers{bg,mybg,main,fg}%
\begin{tikzpicture}[
  remember picture,
  font=\sffamily,
  scale=0.8,
  node distance = 10em,
  boxstyle/.style={rectangle, text width=6.5em, minimum height=1ex, align=center, fill=NavyBlue, draw=white, font=\sffamily\color{white} },
  line/.style={draw, thick, -stealth},
  arrow/.style={thick,->, >=stealth, line cap=round, shorten <=3pt, shorten >=3pt},
  darrow/.style={thick,<->, >=stealth, line cap=round, shorten <=3pt, shorten >=3pt},
  ]

\begin{pgfonlayer}{bg}
\begin{scope}[scale=1.3, local bounding box=DROPLET, shift={(0, -0.33)}]
\draw[thick] (-1.5,-1.5) rectangle (1.5,1.5);
\draw[thick, fill=NavyBlue, fill opacity=0.5] (0,0) circle (1);

\node (GAMMA) at (1.1, 1.2) {$\Gamma(t)$};
\node (INTERFACE1) at (0.7, 0.7) {};
\node (INTERFACE) at (1.0, 0.0) {};
\path[<-, thick] (INTERFACE1) edge [bend right=20] (GAMMA.south);

\node at (0.0, 0.0) {$\cP_{\phasel}$};
\node[right] at (-1.45, -1.2) {$\cP_{\phaser}$};

\node[below, anchor=north, text width=3cm, align=center] (CFDTEXT) at (0, -1.6) {Computational Domain};
\end{scope}
\end{pgfonlayer}

\begin{pgfonlayer}{fg}
\begin{scope}[shift={(4.75, 0.575)}, local bounding box=SI]
  \fill[NavyBlue, opacity=0.5] (-1,-1) rectangle (0.236, 1);
  \draw (-1,-1) rectangle (1,1);
\end{scope}
\node[below, anchor=north, text width=2.5cm, align=center, yshift=-1ex] (SITEXT) at (SI.south) {Sharp interface front capturing};
\end{pgfonlayer}

\begin{pgfonlayer}{mybg}
\node[fill=gray!5, inner sep=2mm, fit=(SI) (SITEXT) ] (SIBOX) {};

\begin{scope}
\node[fit={(INTERFACE.south west) (INTERFACE.north east)}, inner sep=0pt, draw=black, opacity=0.8, thick, fill opacity=0.2, fill=gray] (rect1) {};

\draw[dashed, black, opacity=0.8, shorten <= 0.05cm, shorten >= 0.05cm] (rect1.north west) -- (SI.north west);
\draw[dashed, black, opacity=0.8, shorten <= 0.05cm, shorten >= 0.05cm] (rect1.south west) -- (SI.south west);
\draw[dashed, black, opacity=0.5, shorten <= 0.05cm, shorten >= 0.05cm, thin] (rect1.north east) -- (SI.north east);
\draw[dashed, black, opacity=0.5, shorten <= 0.05cm, shorten >= 0.05cm, thin] (rect1.south east) -- (SI.south east);
\end{scope}
\end{pgfonlayer}

\begin{pgfonlayer}{fg}
\begin{scope}[scale=2.5, shift={(5,-2)},
local bounding box=WAVEPATTERN]

\node[above, align=center, anchor=south] (NUMERICALFLUX) at (0,1.25) {
$g_\phasel(\textcolor{NavyBlue}{\vu_{\phasel}}, \textcolor{NavyBlue}{\vu_{\phaser}})
= f(\textcolor{red!75!black}{\vu^{*}_{\phasel}})
- \textcolor{red!75!black}{s}
\textcolor{red!75!black}{\vu^{*}_{\phasel}}
$
};
\node[above, align=center, anchor=south] (NUMERICALFLUX2) at (NUMERICALFLUX.north) {
$g_\phaser(\textcolor{NavyBlue}{\vu_{\phasel}}, \textcolor{NavyBlue}{\vu_{\phaser}})
= f(\textcolor{red!75!black}{\vu^{*}_{\phaser}})
- \textcolor{red!75!black}{s}
\textcolor{red!75!black}{\vu^{*}_{\phaser}}
$
};

\node[above, align=center, anchor=south] at (NUMERICALFLUX2.north) {Numerical fluxes};
\end{scope}
\end{pgfonlayer}
\node[fill=gray!5, inner sep=2mm, fit=(WAVEPATTERN.south west) (WAVEPATTERN.north east) ] (WAVEPATTERNBOX) {};

\draw (SIBOX.east) edge[out=0,in=180, <->, >=stealth, thick,
line cap=round, shorten <=5pt, shorten >=10pt] node [pos=0.66, below, anchor=north east, text width=2cm, align=center] {} (WAVEPATTERNBOX.west);

\begin{scope}[shift={(3.0,-8.0)}, local bounding box=MD, scale=3]
    \tikzset{arrow/.style={->, >=stealth, line cap=round, shorten <=0pt, shorten >=3pt, opacity=1.0}}

    \node[right, xshift=0mm, anchor=north, text width=4.0cm, align=center] (MDTEXT) at (0.0,-0.25) {Riemann Solver};

    \tikzset{coordinateaxis/.style={->, >=stealth, thick, line join=round, line cap=round}}
\tikzset{shockline/.style={-, line join=round, line cap=round, opacity=0.66}}
\tikzset{rarefactionline/.style={-, line join=round, line cap=round, opacity=0.5}}

\draw[shockline] (0,0) -- ({cos(45)}, {sin(45)});
\draw[shockline, dashed] (0,0) -- ({cos(75)}, {sin(75)}) node[pos=1.0, above right]{$\textcolor{red!75!black}{s}$};

\def \n {7}
\foreach \s in {1,...,\n}
{
	\draw[rarefactionline] (0,0) -- ({cos(120 + 3 * \s)}, {sin(120 + 3 * \s)});
}
\draw[coordinateaxis] (-1,0) -- (1,0) node[pos=1.0, right] {$\spx$}node[pos=0.0, left] {};
\draw[coordinateaxis] (0,0) -- (0,1) node[pos=1.0, above] {$t$};

\node at (-0.7, 0.25) {$\textcolor{NavyBlue}{\vu_{\phasel}}$};
\node at (0.7, 0.25) {$\textcolor{NavyBlue}{\vu_{\phaser}}$};
\node at (-0.20, 0.6) {$\textcolor{red!75!black}{\vu^{*}_{\phasel}}$};
\node at (0.33, 0.6) {$\textcolor{red!75!black}{\vu^{*}_{\phaser}}$};
\end{scope}

\begin{pgfonlayer}{bg}
\node[fill=gray!5, inner sep=2mm, fit=(MD)] (MDBOX) {};
\end{pgfonlayer}

\begin{pgfonlayer}{fg}
\begin{scope}[shift={($(WAVEPATTERNBOX.south)+(0,-3)$)}, %
local bounding box=MODELREDUCTION]

\node[above, align=center, anchor=south] (MRTEXT) at (0,0) {Neural Network};
\end{scope}
\end{pgfonlayer}
\node[fill=gray!5, inner sep=2mm, fit=(MODELREDUCTION.south west) (MODELREDUCTION.north east) ] (MODELREDUCTIONBOX) {};

\draw ([xshift=-2ex] WAVEPATTERNBOX.south) edge[->, bend right=20, >=stealth, thick, NavyBlue,
line cap=round, shorten <=5pt, shorten >=10pt] node [pos=0.5, left, yshift=0ex, text width=2cm, align=center] {Input: $(\textcolor{NavyBlue}{\vu_{\phasel}}, \textcolor{NavyBlue}{\vu_{\phaser}})$} ([xshift=-2ex] MODELREDUCTIONBOX.north);

\draw ([xshift=2ex] WAVEPATTERNBOX.south) edge[<-, bend left=20, >=stealth, thick, red!75!black,
line cap=round, shorten <=5pt, shorten >=10pt] node [pos=0.5, right, yshift=0ex, text width=2cm, align=center] {Response: $(\textcolor{red!75!black}{s}, \textcolor{red!75!black}{\vu^{*}_{\phasel}}, \textcolor{red!75!black}{\vu^{*}_{\phaser}})$} ([xshift=2ex] MODELREDUCTIONBOX.north);

\draw ([xshift=2ex] MDBOX.east) edge[out=0,in=180, ->, >=stealth, thick, black,
line cap=round, shorten <=5pt, shorten >=10pt] node [pos=0.5, above, rotate=50, yshift=0ex, text width=2cm, align=center] {Samples} ([yshift=-2ex] MODELREDUCTIONBOX.west);

\end{tikzpicture}%
}
\caption{Graphical illustration showing how the neural network, the Riemann solver and the front capturing scheme interact.
}
\label{fig:model_overview}
\end{figure}
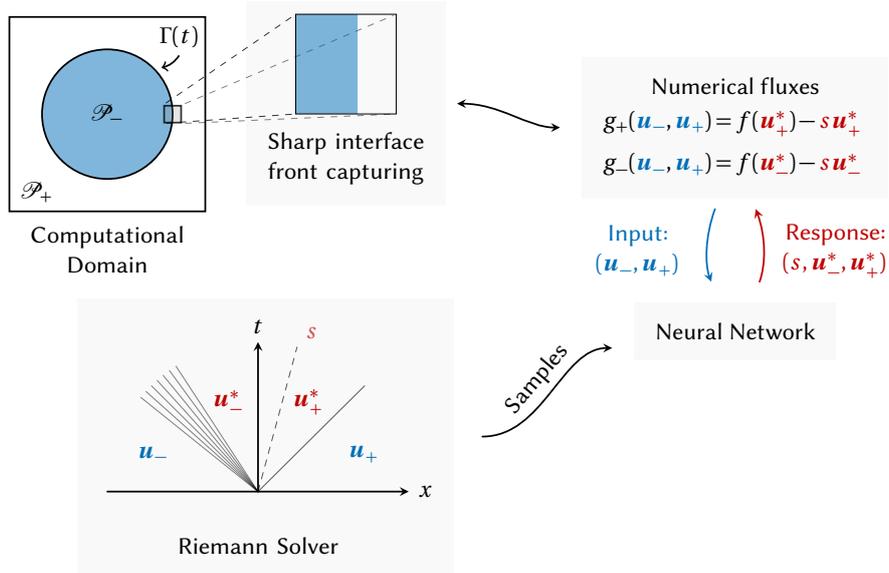

However, such surrogate models are generally not constructed with the intention to satisfy crucial physical constraints like \eqref{eq:rh_condition} even if the original solver satisfies them.
In the present work, we investigate methods for building surrogate models that incorporate knowledge about physical constraints of the underlying system, such as mass or momentum conservation.
To be more specific, we develop methods for building a reliable surrogate Riemann solver, based on neural networks, that incorporates conservation properties such as \eqref{eq:rh_condition} and demonstrate their use in actual numerical simulations.
\newline
Refer to \autoref{fig:model_overview} to get a graphical overview of the interaction between the different algorithmic components.

\section{Neural Networks}
\label{sec:neural_networks}

We are interested in approximating a function of the form
\begin{align} \label{eq:target_function}
F \colon \bR^{d_{\mathrm{in}}} \to \bR^{d_{\mathrm{out}}},
\end{align}
using artificial neural networks. In particular, we consider a specific feed-forward network architecture known as multilayer perceptrons (MLPs), in which the basic computing units (neurons) are stacked in layers. The first layer is called the input layer, and has the sole purpose of providing an input signal to the network. The last layer is the output layer, while all the intermediate layers are known as the hidden layers.

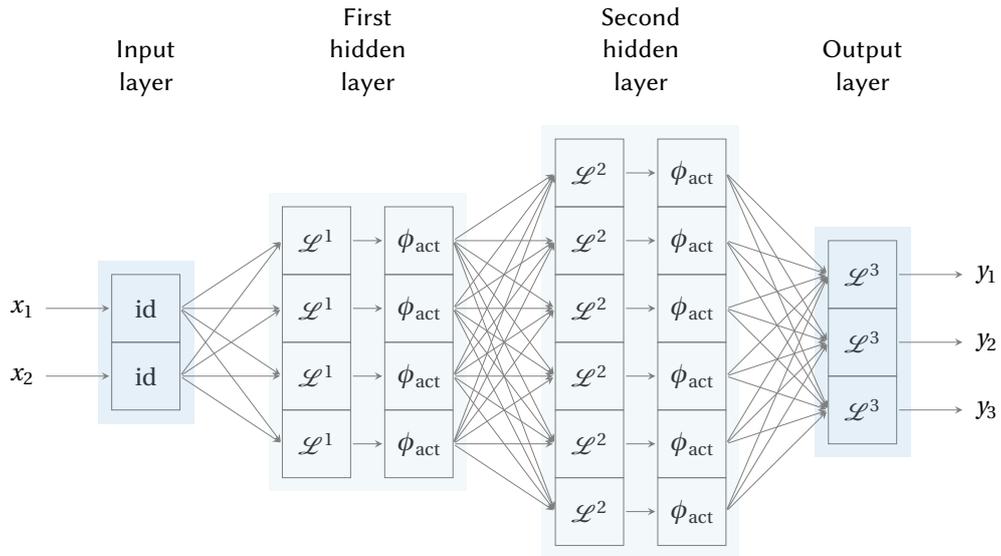
\begin{figure}[ht]
 \centering
\resizebox{0.9\columnwidth}{!}{
\tikzsetnextfilename{feed_forward_network}
\begin{tikzpicture}[shorten >=1pt,->,draw=black!50, node distance=2cm, font=\sffamily]
    \tikzstyle{every pin edge}=[<-,shorten <=1pt]
    \tikzstyle{neuron}=[circle,fill=black!25,minimum size=17pt,inner sep=0pt]
    \tikzstyle{input neuron}=[neuron, fill=NavyBlue!90];
    \tikzstyle{output neuron}=[neuron, fill=NavyBlue!90];
    \tikzstyle{hidden neuron}=[neuron, fill=NavyBlue!50];
    \tikzstyle{annot} = [text width=4em, text centered];
    \tikzstyle{arrow}=[->, draw=black!50, >=stealth, shorten <=1pt, shorten >=1pt];

    \begin{scope}[local bounding box=Ilayer]
    \foreach \name / \y in {1,...,2} {
    \begin{scope}[yshift=-\y cm]
    \draw (0,0) rectangle (1cm,1cm);
    \node[anchor=center, color=black!80] at (0.5cm,0.5cm) {$\mathrm{id}$};
    \coordinate (Ir-\name) at (1cm, 0.5cm);
    \coordinate (Il-\name) at (0cm, 0.5cm);
    \end{scope}
    }
    \end{scope}
    \node[fit=(Ilayer), inner sep=0.2cm, fill=NavyBlue, opacity=0.1] (I-layer) {};

    \foreach \dest in {1,...,2}
            \path ($(Il-\dest)-(1cm,0)$) edge[arrow] node[anchor=east, pos=0] {$x_\dest$} (Il-\dest);

    \begin{scope}[local bounding box=H1layer]
    \foreach \name / \y in {1,...,4} {
    \begin{scope}[xshift=2.5cm, yshift=-\y cm + 1cm]
    \draw (0,0) rectangle (1cm,1cm);
    \node[anchor=center, color=black!80] at (0.5cm,0.5cm) {$\cL^1$};
    \coordinate (H1l-\name) at (0cm, 0.5cm);
    \coordinate (H1r-\name) at (1cm, 0.5cm);
    \end{scope}
    }

    \foreach \name / \y in {1,...,4} {
    \begin{scope}[xshift=4.0cm, yshift=-\y cm + 1cm]
    \draw (0,0) rectangle (1cm,1cm);
    \node[anchor=center, color=black!80] at (0.5cm,0.5cm) {$\phi_{\mathrm{act}}$};
    \coordinate (A1l-\name) at (0cm, 0.5cm);
    \coordinate (A1r-\name) at (1cm, 0.5cm);
    \end{scope}
    \path (H1r-\name) edge[arrow] (A1l-\name);
    }
    \end{scope}
    \node[fit=(H1layer), inner sep=0.2cm, fill=NavyBlue, opacity=0.05] (I-layer) {};

    \foreach \source in {1,...,2}
        \foreach \dest in {1,...,4}
            \path (Ir-\source) edge[arrow] (H1l-\dest);

    \begin{scope}[local bounding box=H2layer]
    \foreach \name / \y in {1,...,6} {
    \begin{scope}[xshift=6.5cm, yshift=-\y cm + 2cm]
    \draw (0,0) rectangle (1cm,1cm);
    \node[anchor=center, color=black!80] at (0.5cm,0.5cm) {$\cL^2$};
    \coordinate (H2l-\name) at (0cm, 0.5cm);
    \coordinate (H2r-\name) at (1cm, 0.5cm);
    \end{scope}
    }

    \foreach \name / \y in {1,...,6} {
    \begin{scope}[xshift=8cm, yshift=-\y cm + 2cm]
    \draw (0,0) rectangle (1cm,1cm);
    \node[anchor=center, color=black!80] at (0.5cm,0.5cm) {$\phi_{\mathrm{act}}$};
    \coordinate (A2l-\name) at (0cm, 0.5cm);
    \coordinate (A2r-\name) at (1cm, 0.5cm);
    \end{scope}
    \path (H2r-\name) edge[arrow] (A2l-\name);
    }
    \end{scope}
    \node[fit=(H2layer), inner sep=0.2cm, fill=NavyBlue, opacity=0.05] (I-layer) {};

    \foreach \source in {1,...,4}
        \foreach \dest in {1,...,6}
            \path (A1r-\source) edge[arrow] (H2l-\dest);

    \begin{scope}[local bounding box=Olayer]
    \foreach \name / \y in {1,...,3} {
    \begin{scope}[xshift=10.5cm, yshift=-\y cm + 0.5cm]
    \draw (0,0) rectangle (1cm,1cm);
    \node[anchor=center, color=black!80] at (0.5cm,0.5cm) {$\cL^3$};
    \coordinate (Ol-\name) at (0cm, 0.5cm);
    \coordinate (Or-\name) at (1cm, 0.5cm);
    \end{scope}
    }
    \end{scope}
    \node[fit=(Olayer), inner sep=0.2cm, fill=NavyBlue, opacity=0.1] (I-layer) {};

    \foreach \source in {1,...,6}
        \foreach \dest in {1,...,3}
            \path (A2r-\source) edge[arrow] (Ol-\dest);

    \foreach \dest in {1,...,3}
         \path (Or-\dest) edge[arrow] node[anchor=west, pos=1] {$y_\dest$} ($(Or-\dest)+(1cm,0)$) ;

    \node[anchor=south, text width=4em, align=center] at (0.5cm, 2.5cm) {Input layer};
    \node[anchor=south, text width=4em, align=center] at (3.75cm, 2.5cm) {First hidden layer};
    \node[anchor=south, text width=4em, align=center] at (7.75cm, 2.5cm) {Second hidden layer};
    \node[anchor=south, text width=4em, align=center] at (11cm, 2.5cm) {Output layer};

\end{tikzpicture}
}
\caption{Sketch of a feed forward neural network with two hidden layers.}
\label{fig:mlp}
\end{figure}

In machine learning terminology, an MLP of depth $K$ corresponds to a network with an input layer, $K-1$ hidden layers and an output layer. Let us denote by $N_l$ the number of neurons in the $l$-th layer, for $l=0,...,K$. Each layer of the network (barring the input layer), receives the output $\vZ_{l-1} \in \bR^{N_{l-1}}$ from the previous layer and performs an affine linear transformation of the form
\begin{align}\label{eq:single_layer}
\cL^l(\vZ^{l-1}) := \vW^l \vZ^{l-1} + \vb^l,
\end{align}
where $\vW^l$ and $\vb^l$ are respectively the weights and biases associated with the layer $l$. The transformed vector is then acted upon (component-wise) by a non-linear activation function $\phi_{\mathrm{act}}$ to form the input vector $\vZ^l$ for the next layer. The activation function prevents the neural network from collapsing into a single affine linear transformation. Several choices for $\phi_{\mathrm{act}}$ have been proposed, each with its own advantages \cite{glorot.bordes.ea:deep:2011,goodfellow.bengio.ea:deep:2016,ramachandran.zoph.ea:searching:2017}. In this work, we use the ELU activation function \cite{clevert.unterthiner.ea:fast:2015}
\begin{align*}
 \phi_{\mathrm{act}}^{\mathrm{elu}}(z) =
 \begin{cases}
 \exp(z) - 1 & \text{ for } z < 0, \\
 z & \text{ for } z \geq 0.
 \end{cases}
\end{align*}
Furthermore, we set the activation function after the output layer to be the identity function. To ensure consistency, we set $N_0 = d_{\mathrm{in}}$, $N_K = d_{\mathrm{out}}$ and $\vZ_0 = \ix$. Thus, we have the neural network representation
\begin{align} \label{eq:neural_net_function}
F_{\vec{\theta}}(\ix)  = \bigl(\cL^K  \circ \phi_{\mathrm{act}} \circ \cL^{K-1} \circ ... \circ \phi_{\mathrm{act}} \circ \cL^1\bigr)(\ix),
\end{align}
where $\vec{\theta} = \{\vW^l,\vb^l \}_{l=1}^K$ is the set of trainable parameters of the network.  The schematic of an MLP with depth 2 is shown in \autoref{fig:mlp}.

The \textit{training} of the neural network involves finding the parameters $\vec{\theta}$ to suitably approximate \eqref{eq:target_function}. In the framework of supervised learning, this is achieved by choosing a set of labeled data
\[
D = \bigl\{(\ix_i,\oy_i) \ : \ \oy_i = F(\ix_i) \bigr\},
\]
and choosing a suitable loss function to measure the discrepancy between the true label and network output.
A popular choice for problems of function regression is the mean-squared error loss function
\begin{align} \label{eq:mse_loss}
 L_{\mathrm{mse}}(F_{\vec{\theta}}, D) = \frac{1}{\abs{D}} \sum_{(\ix,\oy) \in D} \norm*{\oy - F_{\vec{\theta}}(\ix)}^2.
\end{align}
The \textit{optimal} value of $\vec{\theta}$ minimizes \eqref{eq:mse_loss}.

The non-linear and non-convex dependency of the loss-function with respect to $\vec{\theta}$ makes the determination of an exact optimal $\vec{\theta}$ intractable.
Thus, it is customary to use an iterative algorithm, such as stochastic gradient descent, to obtain an approximation to this value.
Each step of the optimization scheme is usually performed for several data points --- a so-called \emph{batch} --- at once to approximate the gradients.
In this context, one iteration over the whole training data set $D_{\mathrm{train}}$ is called an \emph{epoch}.
We refer the interested readers to \cite{goodfellow.bengio.ea:deep:2016,ruder:overview:2016} for a detailed discussion on the various iterative optimizers used in practice.

A typical problem faced while training neural networks involves overfitting the training data.
This can severely deteriorate the networks capacity to generalize, i.e., accurately predict the output for data points not utilized while training.
One strategy to circumvent this issue involves the manual termination of the iterative optimization process based on cross-validation.
More precisely, the data set $D$ is split into a training set $D_{\mathrm{train}}$ and a validation set $D_{\mathrm{val}}$.
The network is trained to minimize the loss only over $D_{\mathrm{train}}$, while the loss is evaluated on the $D_{\mathrm{val}}$ at the end of each epoch as an indicator of the generalization error.
The training is terminated if the validation loss does not improve for a pre-decided number of epochs, or the maximum number of epochs is reached.

An alternate strategy to avoid over-fitting involves augmenting the loss-function \eqref{eq:mse_loss} with a penalty term on the weights $\vW$ of the network
\begin{align} \label{eq:weight_decay}
 P_{\mathrm{wd}}(\vec{\theta}) \coloneqq \alpha_{\mathrm{wd}} \sum_{l=1}^K \abs{\vW^l}^p, \end{align}
where $\alpha_{\mathrm{wd}} \geq 0$ is a tunable hyper-parameter that controls the amount of weight decay.
Typically, one chooses $p=1$ (to induce sparsity) or $p=2$ as the exponent in \eqref{eq:weight_decay}.
In this work, weight decay is observed for all layers except the output layer.
Other methods to avoid over-fitting are described in \cite{goodfellow.bengio.ea:deep:2016}.

\begin{remark}[Loss scaling]
In many machine learning applications it is important to normalize the output labels, as well as the input features.
However, as we are interested in fulfilling constraints, label scaling would introduce other technical inconveniences.
We therefore employ an adapted mean squared loss function for the training of the neural networks.
Thus, we first compute the vector of reciprocal empirical standard deviations $\vec{\sigma}_{D}^{-1} = (\tfrac{1}{\sigma_{1,D}}, \ldots , \tfrac{1}{\sigma_{d_{\mathrm{out}},D}})$ of the labels $(\oy_i)_{i=1, \ldots,N}$ in the data set $D$.
The scaled mean squared loss takes the form
\begin{align}
 L_{\mathrm{smse}}(F_{\vec{\theta}}, D) = \frac{1}{\abs{D}} \sum_{(\ix,\oy) \in D}  \norm[\big]{\vec{\sigma}_{D}^{-1} \cdot (\oy - F_{\vec{\theta}}(\ix))}^2.
\end{align}
This ensures that that an equal importance is given to the error corresponding to each component of the output vector.
\newline
In the following, we perform input feature normalization and loss scaling without explicitly referring to it.
\end{remark}

\section{Constraint-Aware Learning}
\label{sec:constraint-aware-methods}

To construct a machine-learned surrogate Riemann solver, as described in \autoref{sec:problem_description}, we seek to train a neural network that approximates the function $\cR$ in \eqref{eq:riemannsolver:generic} with input values
\begin{align} \label{eq:generic_input}
 \ix \coloneqq (\vu_{\phasel}, \vu_{\phaser}) \in \cP_{\phasel} \times \cP_{\phaser} \eqqcolon \cX
\end{align}
and output values
\begin{align} \label{eq:generic_output}
 \oy \coloneqq (\vu^{*}_{\phasel}, \vu^{*}_{\phaser}, s) \in \bR^m \times \bR^m \times \bR \eqqcolon \cY.
\end{align}
Our goal is to build a surrogate model $F_{\vec{\theta}}$ for $\cR$ that is constraint-aware, i.e. during the construction of $F_{\vec{\theta}}$ (physical) constraints are taken into account.
\newline
In this work we restrict ourselves to algebraic constraints
 described by the function $\Phi \colon \cX \times \cY \to \bR^+$.
An input--output tuple $(\ix,\oy) \in \cX \times \cY$ is said to fulfil the constraint $\Phi$ if we have
\begin{align} \label{eq:generic_constraint}
 \Phi(\ix, \oy) = 0.
\end{align}
Constraints of this form can describe the underlying physical constraints of the model, such as mass conservation.
More specifically, the Rankine--Hugoniot condition \eqref{eq:rh_condition} can be written in the form \eqref{eq:generic_constraint} using the constraint function
\begin{align} \label{eq:constraint:rh}
 \Phi_{\mathrm{RH}}(\ix, \oy)
 =  \Phi_{\mathrm{RH}}\bigl((\vu_{\phasel}, \vu_{\phaser}), (\vu^{*}_{\phasel}, \vu^{*}_{\phaser}, s)\bigr)
 \coloneqq \norm*{s \, (\vu^{*}_{\phasel} - \vu^{*}_{\phaser}) - \bigl(f(\vu^{*}_{\phasel}) - f(\vu^{*}_{\phaser})\bigr) }_1.
\end{align}
A surrogate model $F_{\vec{\theta}}$, as in \eqref{eq:neural_net_function}, fulfills a constraint exactly if
\begin{align}
 \Phi\bigl(\ix, F_{\vec{\theta}}(\ix) \bigr) = 0  \quad \text{ for all } \ix \in \cX.
\end{align}
Then, an ideal surrogate model $F_{\vec{\theta}}$ would be a solution of the optimization problem
\begin{align} \label{eq:ideal_optimization_problem}
 \text{minimize } & L_{\mathrm{mse}}(F_{\vec{\theta}}, D) ~\text{ for all } D \subset \cX \times \cY \\
 \text{subject to } & \Phi\bigl(\ix, F_{\vec{\theta}}(\ix) \bigr) = 0 ~\text{ for all } x \in \cX.
\end{align}
This is however often a strong condition on $F_{\vec{\theta}}$, due to the complexity of the constraint function $\Phi$.
In practice it might suffice to keep the constraint deviation small, i.e. by considering a relaxed version of the optimization problem \eqref{eq:ideal_optimization_problem} for a tolerance $\varepsilon > 0$, i.e.
\begin{align} \label{eq:relaxed_optimization_problem}
 \text{minimize } & L_{\mathrm{mse}}(F_{\vec{\theta}}, D) ~\text{ for all } D \subset \cX \times \cY \\
 \text{subject to } & \bigl\lvert\Phi\bigl(\ix, F_{\vec{\theta}}(\ix) \bigr) \bigr\rvert \leq \varepsilon ~\text{ for all } \ix \in \cX.
\end{align}
In the next two subsections we present two approaches to build constraint-aware surrogate models.
The first approach satisfies constraints exactly and corresponds to the optimization problem \eqref{eq:ideal_optimization_problem}.
The second approach is a relaxed method which only penalizes deviations from the constraints, and is an example of optimization problem \eqref{eq:relaxed_optimization_problem}, albeit without giving explicit estimates for the tolerance parameter $\varepsilon$.

\subsection{Constraint-resolving Layer Method (\cres)}
\label{sec:constraint_layer_net}

If it is possible to resolve the constraint equation \eqref{eq:generic_constraint} with respect to e.g. components of the output data, one can infer even more knowledge about the system.
Let us assume that the output variable $\oy$ can be written as a vector of the form $\oy \cong (\vec{z},\vec{q}) \in \cZ \times \cQ \cong \cY$ such that there exists a function
$\Psi \colon \cX \times \cZ \to \cQ$ fulfilling
\begin{align} \label{eq:psi}
 \Phi\bigl(\ix, (\vec{z}, \Psi(\ix,\vec{z}))\bigr) = 0 \quad \text{ for all } \ix \in \cX, \vec{z} \in \cZ.
\end{align}

\begin{figure}[ht]
 \centering
\resizebox{0.7\columnwidth}{!}{
\tikzsetnextfilename{resolving_network}
\begin{tikzpicture}[draw=black!50, node distance=2cm, font=\sffamily]
    \tikzstyle{every pin edge}=[<-,shorten <=1pt, shorten >=1pt]
    \tikzstyle{connection}=[->,shorten <=1pt, shorten >=1pt]
    \tikzstyle{neuron}=[circle,fill=black!25,minimum size=17pt,inner sep=0pt]
    \tikzstyle{input neuron}=[neuron, fill=NavyBlue!90];
    \tikzstyle{output neuron}=[neuron, fill=NavyBlue!90];
    \tikzstyle{hidden neuron}=[neuron, fill=NavyBlue!50];
    \tikzstyle{annot} = [text width=4em, text centered]

    \foreach \name / \y in {1,...,2}
        \node[input neuron, pin=left:$\ix_\y$] (I-\name) at (0,-\y) {};

    \foreach \name / \y in {1,...,5}
        \path[yshift=1.5cm]
            node[hidden neuron] (H1-\name) at (2cm,-\y cm) {};

    \foreach \name / \y in {1,...,5}
        \path[yshift=1.5cm]
            node[hidden neuron] (H2-\name) at (4cm,-\y cm) {};

    \foreach \name / \y in {1,...,2}
        \path[yshift=0cm]
            node[output neuron] (O-\name) at (6cm,-\y cm) {};

    \path[yshift=0cm]
            node[hidden neuron] (TMPVAR) at (6cm,0cm) {};

    \foreach \source in {1,...,5}
        \foreach \dest in {1,...,5}
            \path (H1-\source) edge[connection, opacity=0.5, densely dashed] (H2-\dest);

    \foreach \source in {1,...,2}
        \foreach \dest in {1,...,5}
            \path (I-\source) edge[connection] (H1-\dest);

    \foreach \source in {1,...,5}{
        \path (H2-\source) edge[connection] (TMPVAR);
        \foreach \dest in {1,...,2}
            \path (H2-\source) edge[connection] (O-\dest);
    }

    \node[right of=O-1, xshift=1.8cm] (Olabel-1) {$\oy_2$};
    \node[right of=O-2, xshift=1.8cm] (Olabel-2) {$\oy_3$};
    \foreach \source in {1,...,2}
            \path (O-\source) edge[connection] (Olabel-\source);

    \node[above of=Olabel-1, node distance=1cm] (Olabel-3) {$\oy_1$};

    \node[neuron, fill=red!50, text=red!75!black, left of=Olabel-3, xshift=0.25cm] (O-con)  {$\Psi$};
    \foreach \source in {1,...,2}
            \path (O-\source) edge[connection, color=red!50, thick] (O-con);
    \path (TMPVAR) edge[connection, color=red!50, thick] (O-con);

    \path (O-con) edge[connection, color=red!50, thick] (Olabel-3);

    \begin{scope}[color=red!50, thick, opacity=0.9]
    \node (C1) at (2cm,1.2cm) {};
    \node (C2) at (6cm,1.2cm) {};
    \draw[shorten <=1pt] (I-1) to[out=0,in=180] (C1.center);
    \draw (C1.center) to[out=0,in=180] (C2.center);
    \draw[->, shorten >=1pt] (C2.center) to[out=0, in=150, looseness=1] (O-con);
    \draw[shorten <=1pt] (I-2) to[out=0,in=180, looseness=0.9] ($(C1.center) - (0,2pt)$);
    \draw ($(C1.center) - (0,2pt)$) to[out=0,in=180] ($(C2.center) - (0,2pt)$);
    \draw[->, shorten >=1pt] ($(C2.center) - (0,2pt)$) to[out=0, in=160, looseness=1] (O-con);
    \end{scope}

    \node[annot, above, node distance=2cm, yshift=1.0cm] (hl) at ($(H1-1)!0.5!(H2-1)$) {Hidden layers};
    \node[annot,left of=hl, node distance=3cm] {Input layer};
    \node[annot,right of=hl, node distance=3cm] (Olayerlabel) {Output layer};
    \node[annot, text width=5em, right of=Olayerlabel, node distance=2cm] {Constraint layer};
\end{tikzpicture}
}
\caption{A sketch of the structure of a neural network that resolves the constraint analytically.}
\label{fig:resolving_layer}
\end{figure}
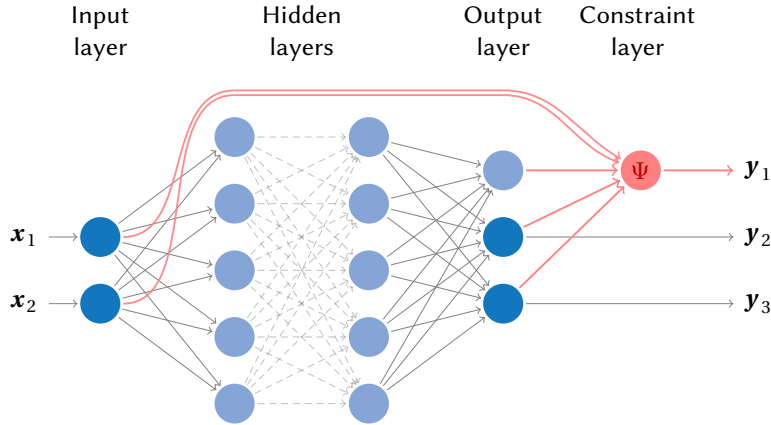

Using the function $\Psi$, we can add a new layer to the network that performs the mapping defined by $\Psi$ --- see \autoref{fig:resolving_layer}.
More formally, the neural network mapping $F_{\vec{\theta}}$, as in \eqref{eq:neural_net_function}, is composed of a sub-network $G_{\vec{\theta}} \colon \cX \to \cZ$ and $\Psi$, i.e.
\begin{align} \label{eq:resolving_network_function}
 F_{\vec{\theta}}(\ix) \coloneqq \bigl( G_{\vec{\theta}}(\ix), \Psi(\ix, G_{\vec{\theta}}(\ix)) \bigr).
\end{align}
Hence, $\Phi(\ix, F_{\vec{\theta}}(\ix)) = 0$ for all $\ix \in \cX$, and $F_{\vec{\theta}}$ resolves the constraint exactly.
Accordingly, we call this construction the \emph{constraint-resolving layer method (\cres)}, if $F_{\vec{\theta}}$ is of the form \eqref{eq:resolving_network_function}.
We note that with the output layer depending on the input $x$, we have a slightly modified version of MLPs compared to \autoref{sec:neural_networks}.
For explicit examples in the framework of hyperbolic conservation laws and Rankine--Hugoniot conditions we refer to \autoref{sec:models}.

The advantage of the constraint-resolving layer method, compared to simply computing $\vec{q} = \Psi(\ix,\vec{z})$ as a post-processing step, is that we can use all information from the data set during the training process, while incorporating further knowledge about the system. 
The drawback, however, is that the constraint \eqref{eq:generic_constraint} has to be handled analytically and must be uniquely solvable.
Therefore a deeper understanding of the model and constraint is needed.

If the constraint can not be directly expressed, condition \eqref{eq:psi} on $\Psi$ can be loosened, by demanding that $\Psi$ solves \eqref{eq:psi} only approximately --- see \autoref{sec:iso_euler} for an example where $\Psi$ is expressing the constraint \eqref{eq:generic_constraint} approximately.

\subsection{Constraint-adapted Loss Method (\cadl)}
\label{sec:constraint_loss_net}
The second, more general method to build constraint-aware neural networks is based on the idea of penalizing deviations from the constraint \eqref{eq:generic_constraint} during the training procedure (as mentioned in \cite{marquez-neila.salzmann.ea:imposing:2017, lee.wick.ea:gradient:2017}) .
To this end, we introduce the constraint loss function
\begin{align} \label{eq:constraint_loss}
  L_{\Phi}(F_{\vec{\theta}}, D) = \frac{1}{\abs{D}} \sum_{(\ix,\:\cdot\:) \in D}  \abs[\big]{\Phi(\ix, F_{\vec{\theta}}(\ix))}.
\end{align}
Obviously, $L_{\Phi}(F_{\vec{\theta}}, D) = 0$ holds if $F_{\vec{\theta}}$ fulfills the constraint $\Phi(\ix, F_{\vec{\theta}}(\ix)) = 0$ on the data set $D$.
Combined with the mean-squared error loss we obtain the constraint-adapted loss function
\begin{align} \label{eq:constraint_adapted_loss}
  L_{\mathrm{mse},\Phi}(F_{\vec{\theta}}, D) = \frac{1}{\abs{D}} \sum_{(\ix,\oy) \in D} \biggl(  \norm[\big]{\oy - F_{\vec{\theta}}(\ix)}^2  + \lambda \abs[\big]{\Phi(\ix, F_{\vec{\theta}}(\ix))} \biggr),
\end{align}
where $\lambda > 0$ is an adjustable constraint penalty parameter.
\newline
Neural networks that are trained with the loss function \eqref{eq:constraint_adapted_loss} are said to be generated with the \emph{constraint-adapted loss method (\cadl)}.
The advantage of building surrogate models with this method is that it is straightforward to apply and independent of the underlying model, i.e., no deeper knowledge about the constraint is needed (as opposed to the \cres-method).
The disadvantage, however, is that it only considers the constraint in a relaxed manner, i.e. the deviation from the constraint is reduced but the constraint is not guaranteed to be satisfied exactly.

\section{Case Study Model Problems}
\label{sec:models}
In this section we introduce three different models to which we apply the constraint-aware learning methods (\cres) and (\cadl) from \autoref{sec:constraint-aware-methods}.
\newline
The first model problem is a scalar conservation law with a cubic flux, which will serve as a toy problem to test the proposed constraint-aware methods.
The next two models are more realistic test cases and comprise the isothermal Euler equations for two-phase flow and the Euler equations for an ideal gas.

Although the constraint-aware learning methods can be applied for constraints depending on both the input and output variables $\ix$ and $\oy$, the model constraints considered below are of the form $\Phi = \Phi(\oy)$.

\subsection{Cubic Flux Model: Undercompressive Wave}
\label{sec:cubic}

\begin{figure}[ht]
 \centering
\resizebox{0.45\columnwidth}{!}{
\tikzsetnextfilename{wave_pattern_cubic}
\begin{tikzpicture}[scale=3.0, font=\sffamily]

\tikzset{coordinateaxis/.style={->, >=stealth, thick, line join=round, line cap=round}}
\tikzset{shockline/.style={-, line join=round, line cap=round, opacity=1.0}}
\tikzset{rarefactionline/.style={-, line join=round, line cap=round, opacity=0.66}}

\draw[shockline, NavyBlue, thick, densely dashed] (0,0) -- ({cos(75)}, {sin(75)}) node[pos=1.0, above right]{$s$};

\def \n {5}
\foreach \s in {1,...,\n}
{
	\draw[rarefactionline] (0,0) -- ({cos(30 + 3 * \s)}, {sin(30 + 3 * \s)});
}

\draw[coordinateaxis] (-1,0) -- (1,0) node[pos=1.0, right] {$\spx$};
\draw[coordinateaxis] (0,0) -- (0,1) node[pos=1.0, above] {$t$};

\node[NavyBlue, anchor=east] at (-0.2, 0.25) {$\textcolor{black}{u_{\phasel}} = u^{*}_{\phasel} \in \cP_\phasel $};
\node[NavyBlue, rotate=60] at (0.375, 0.65) {$u^{*}_{\phaser} \in \cP_\phaser$};
\node[anchor=west] at (0.6, 0.25) {$u_{\phaser}$};

\end{tikzpicture}
}
\caption{Example of a Riemann solution consisting of a rarefaction wave followed by an undercompressive shock wave. The wave of interest is the undercompressive wave.}
\label{fig:wave_pattern:cubic}
\end{figure}
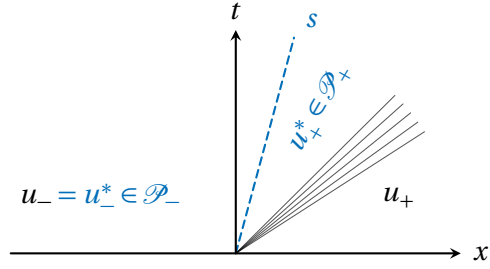

In this section we consider the cubic flux model
\begin{align} \label{eq:cubic_cl}
\partial_t u + \partial_\spx (u^3) = 0
\quad \Leftrightarrow \quad
\partial_t u + \partial_\spx f(u) = 0,
\end{align}
for $u \colon \bR \times [0, t_{\mathrm{end}}) \to \cU$ with $\cU = \bR$.
This model can be seen as a simplified prototype model for phase transition dynamics \cite{chalons.lefloch:computing:2003}.
\newline
As the flux is non-convex and has an inflection point at $u = 0$ (undercompressive) shock waves may occur \cite{lefloch:hyperbolic:2002}, and can be interpreted as phase boundaries.
These waves can only occur at jumps between the two phases  $\cP_{\phasel} \coloneqq \{u > 0\}$ and $\cP_{\phaser} \coloneqq \{u < 0\}$.
\newline
The solution to the Riemann problem \eqref{eq:riemannproblem} for \eqref{eq:cubic_cl}, with $u_{\phaselr} \in \cP_\phaselr$ may consist of two elementary waves.
One of these waves is an undercompressive wave jumping from $u^{*}_{\phasel} \in \cP_\phasel$ to $u^{*}_{\phaser} \in \cP_\phaser$, which is the \emph{wave of interest} for this particular model problem --- see \autoref{fig:wave_pattern:cubic} for a depiction of an exemplary wave pattern.
\newline
The uniqueness of the Riemann solution is ensured by requiring that the solution obeys an additional algebraic relation, the so-called kinetic relation \cite{hayes.lefloch:non:1997}.
A particularly simple choice for a kinetic relation is $\varphi(u) = 0$ with
\begin{align}\label{eq:kinrel}
 \varphi \colon \cP_{\phasel} \cup \{0\} \to \cP_{\phaser} \cup \{0\}
 : u \mapsto - \kappa u,
\end{align}
with $\kappa \in [0.5, 1]$.
To solve the Riemann problem \eqref{eq:riemannproblem} with $u_{\phaselr} \in \cP_\phaselr$ and to obtain the corresponding states $(u^{*}_{\phasel}, u^{*}_{\phaser}, s)$ for the wave of interest, we employ the exact Riemann solver described in \cite{chalons.engel.ea:conservative:2014} --- see \autoref{sec:cubic_riemannsolver}.
As in \eqref{eq:riemannsolver:generic}, the Riemann solver defines a function $\cR_{\mathrm{cubic}} \colon \cP_{\phasel} \times \cP_{\phaser} \to  \cP_{\phasel} \times \cP_{\phaser} \times \bR$:
\begin{align} \label{eq:riemannsolver:cubic}
\begin{aligned}
 \cR_{\mathrm{cubic}}(u_{\phasel}, u_{\phaser})
 &= (u^{*}_{\phasel}, u^{*}_{\phaser}, s) \\
  &= \begin{cases}
  \bigl(u_{\phasel},~u_{\phaser},~\overline{s}(u_{\phasel}, u_{\phaser}) \bigr)
  & \text{ for } u_{\phaser} \in \bigl[(\kappa - 1) u_{\phasel}, 0\bigr), \\
  \bigl(u_{\phasel},~\varphi(u_{\phasel}),~\overline{s}(u_{\phasel}, \varphi(u_{\phasel}) )  \bigr)
  & \text{ for } u_{\phaser} < (\kappa - 1)u_{\phasel},
 \end{cases}
\end{aligned}
\end{align}
where $s$ is the wave speed and $u^{*}_{\phasel}$, $u^{*}_{\phaser}$ the adjacent states of the undercompressive wave when $u_{\phasel} \in \cP_{\phasel}$ and $u_{\phaser} \in \cP_{\phaser}$.
Furthermore, $\overline{s}(u_{\phasel}, u_{\phaser})$ is defined as
\begin{align} \label{eq:scalar_wave_speed}
 \overline{s}(u_{\phaser}, u_{\phasel}) \coloneqq
 \begin{cases}
 \dfrac{f(u_{\phaser}) - f(u_{\phasel})}{u_{\phaser} - u_{\phasel}} & \text{ for } u_{\phaser} \neq u_{\phasel}, \\
 f'(u_{\phaser}) & \text{ if } u_{\phaser} = u_{\phasel}.
 \end{cases}
\end{align}
The wave defined by $(u^{*}_{\phasel}, u^{*}_{\phaser}, s)$ fulfills the Rankine--Hugoniot condition
\begin{align} \label{eq:rh:cubic}
  s \, (u^{*}_{\phasel} - u^{*}_{\phaser}) =  f(u^{*}_{\phasel}) - f(u^{*}_{\phaser}),
\end{align}
which will be the constraint that a surrogate model for \eqref{eq:riemannsolver:cubic} should satisfy.
As in \eqref{eq:constraint:rh} the constraint target function $\Phi_{\mathrm{cubic}} \colon \bR \times \bR \times \bR$, corresponding to \eqref{eq:rh:cubic}, is given by
\begin{align} \label{eq:constraint:cubic}
 \Phi_{\mathrm{cubic}}(u^{*}_{\phasel}, u^{*}_{\phaser}, s) \coloneqq \abs[\big]{ s \cdot (u^{*}_{\phasel} - u^{*}_{\phaser}) - (f(u^{*}_{\phasel}) - f(u^{*}_{\phaser})) } = 0.
\end{align}
Our goal is to learn a model to approximate the Riemann solver $\cR_{\mathrm{cubic}}$ while trying to uphold the constraint $\Phi_{\mathrm{cubic}}(u^{*}_{\phasel}, u^{*}_{\phaser}, s) = 0$.
We are in a situation where the mapping $\cR_{\mathrm{cubic}}$ respects the constraint \eqref{eq:constraint:cubic} and therefore we have to ensure that the constraint surrogate model satisfies it as well.

As \eqref{eq:cubic_cl} is a scalar  law, we can resolve the constraint \eqref{eq:constraint:cubic} by computing the correct wave speed $s$ for a jump from $u^{*}_{\phasel}$ to $u^{*}_{\phaser}$ from \eqref{eq:rh_condition}.
The resolving function $\Psi$ (see \autoref{sec:constraint_layer_net}) is therefore given by
\begin{align} \label{eq:psi:cubic}
\Psi_{\mathrm{cubic}}(u^{*}_{\phaser}, u^{*}_{\phasel}) = 
\overline{s}(u^{*}_{\phasel}, u^{*}_{\phaser}) \coloneqq 
 \begin{cases}
 \dfrac{f(u^{*}_{\phasel}) - f(u^{*}_{\phaser})}{u^{*}_{\phasel} - u^{*}_{\phaser}} & \text{ for } u^{*}_{\phaser} \neq u^{*}_{\phasel}, \\ 
 \hfill f'(u^{*}_{\phaser}) & \text{ if } u^{*}_{\phaser} = u^{*}_{\phasel}.
 \end{cases} 
\end{align}

\subsection{Isothermal Two-Phase Flow: Phase Boundary}
\label{sec:iso_euler}

\begin{figure}[ht]
 \centering
\resizebox{0.45\columnwidth}{!}{
\tikzsetnextfilename{vdw_pressure}
\begin{tikzpicture}[scale=1.0, font=\sffamily]
    \begin{axis}[domain=0:2.5, ymin=0.0, ymax=1.5, legend style={at={(0.05,0.95)},anchor=north west}, xlabel=$\rho$, ylabel=$p(\rho)$, grid=both, samples=200, minor x tick num=1, minor y tick num=2, xmin=0, xmax=2.25] %

    \addplot[name path=fcrit, NavyBlue, dashed, very thick] {(8/3) * x * 1.0 / (1 - (1/3) * x) - 3 * x * x};
    \addplot[name path=f, NavyBlue, very thick] {(8/3) * x * 0.85 / (1 - (1/3) * x) - 3 * x * x};

    \legend{$T_{\mathrm{ref}}=T_{\mathrm{crit}}$, $T_{\mathrm{ref}} < T_{\mathrm{crit}}$}
    \end{axis}
\end{tikzpicture}%
}
\caption{Graph of the van der Waals pressure.}
\label{fig:vdw_pressure}
\end{figure}
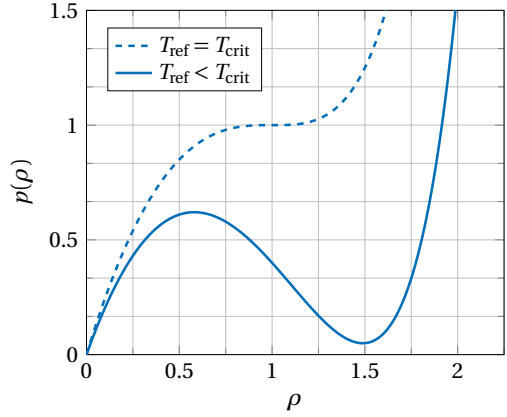

We consider the dynamics of a compressible, inviscid, homogeneous fluid that occurs in liquid or vapor phase.
At constant temperature and in one spatial dimension, the fluid flow can be described by the isothermal Euler equations
\begin{align} \label{eq:iso_euler}
\partial_t \begin{pmatrix} \rho \\ m \end{pmatrix}
+ \partial_\spx \begin{pmatrix} m \\ \tfrac{1}{\rho} m^2 + p(\rho) \end{pmatrix}
= \vec{0}
\quad \Leftrightarrow \quad
\partial_t \vu + \partial_\spx f(\vu) = \vec{0},
\end{align}
where the unknowns $\rho$ and $m$ denote the fluid density and the fluid momentum.
In the context of this model we will write $\vu = (\rho, m)$ for the conservative variables.
To close \eqref{eq:iso_euler} we consider the van der Waals equation of state
\begin{align} \label{eq:vdw_pressure}
p(\rho) = \frac{\rho RT_{\mathrm{ref}}}{1 - b \rho} - a \rho^2,
\end{align}
with the specific gas constant $R = \tfrac{8}{3}$, the reference temperature $T_{\mathrm{ref}} = 0.85$ and parameters $a = 3$ and $b = \tfrac{1}{3}$.
For this choice the critical temperature is computed as $T_{\mathrm{crit}} = 1$. For any reference temperature below the critical temperature the van der Waals pressure becomes non-monotone, see \autoref{fig:vdw_pressure}.
We denote the interval where the pressure decreases, i.e. $p'(\rho) < 0$ for all $\rho \in \cA_{\mathrm{spin}}$, by $\cA_{\mathrm{spin}} \coloneqq (\rho^{\mathrm{max}}_{\phaser}, \rho^{\mathrm{min}}_{\phasel})$.
In this so-called spinodal region $\cA_{\mathrm{spin}}$ the system changes its character and becomes elliptic instead of hyperbolic.
Thus, we want to avoid values in the spinodal region and define the admissible set of densities by $\cA_{\mathrm{2p}} \coloneqq (0,b) \setminus (\rho^{\mathrm{max}}_{\phaser}, \rho^{\mathrm{min}}_{\phasel})$.
The admissible set $\cA_{\mathrm{2p}}$, where the system is hyperbolic, can be split into the liquid phase $\cA_\phasel \coloneqq (\rho^{\mathrm{min}}_{\phasel}, b)$ and the vapor phase $\cA_\phaser \coloneqq (0, \rho^{\mathrm{max}}_{\phaser})$.
Thus, the corresponding domains in the state space are defined by $\cP_\phasel \coloneqq \cA_\phasel \times \bR$ and $\cP_\phaser \coloneqq \cA_\phaser \times \bR$.

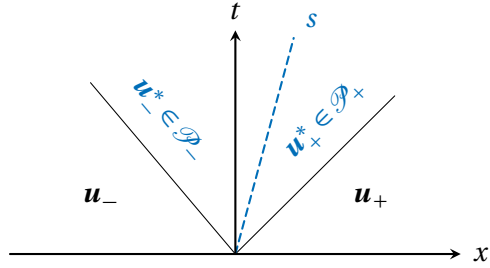
\begin{figure}[ht]
 \centering
\resizebox{0.45\columnwidth}{!}{
\tikzsetnextfilename{wave_pattern_iso}
\begin{tikzpicture}[scale=3.0, font=\sffamily]

\tikzset{coordinateaxis/.style={->, >=stealth, thick, line join=round, line cap=round}}
\tikzset{shockline/.style={-, line join=round, line cap=round, opacity=1.0}}
\tikzset{rarefactionline/.style={-, line join=round, line cap=round, opacity=0.66}}

\draw[shockline] (0,0) -- ({cos(45)}, {sin(45)});
\draw[shockline, NavyBlue, thick, densely dashed] (0,0) -- ({cos(75)}, {sin(75)}) node[pos=1.0, above right]{$s$};
\draw[shockline] (0,0) -- ({cos(130)}, {sin(130)});

\draw[coordinateaxis] (-1,0) -- (1,0) node[pos=1.0, right] {$\spx$};
\draw[coordinateaxis] (0,0) -- (0,1) node[pos=1.0, above] {$t$};

\node at (-0.6, 0.25) {$\vu_{\phasel}$};
\node at (0.6, 0.25) {$\vu_{\phaser}$};
\node[NavyBlue, rotate=-50] at (-0.3, 0.6) {$\vu^{*}_{\phasel} \in \cP_\phasel$};
\node[NavyBlue, rotate=45] at (0.4, 0.6) {$\vu^{*}_{\phaser} \in \cP_\phaser$};

\end{tikzpicture}
}
\caption{Example of a Riemann solution consisting of three shock waves. The wave of interest is the nonclassical shock wave representing the phase boundary.}
\label{fig:wave_pattern:iso}
\end{figure}

We consider the Riemann problem \eqref{eq:riemannproblem}
with $\vu_{\phasel} \in \cP_\phasel$ and $\vu_{\phaser} \in \cP_\phaser$.
In this context the \emph{wave of interest} is the (discontinuous) elementary wave that jumps from a state $\vu^{*}_{\phasel} \in \cP_\phasel$, to $\vu^{*}_{\phaser} \in \cP_\phaser$, i.e. the boundary between the two phases.
A typical Riemann solution of \eqref{eq:riemannproblem} is sketched in \autoref{fig:wave_pattern:iso} .

The Rankine--Hugoniot conditions \eqref{eq:rh_condition} for the system \eqref{eq:iso_euler} are of the form
\begin{align} \label{eq:rh:iso}
\begin{aligned}
  \jump[\big]{ m - s \rho } &= 0, \\
  \jump[\big]{\tfrac{1}{\rho} m \cdot m + p - s m } &= 0,
\end{aligned}
\end{align}
with $\jump{\:\cdot\:} \coloneqq [\:\cdot\:]_\phasel - [\:\cdot\:]_\phaser$ denoting the jump operator.
To obtain unique weak solution we have to impose an additional entropy criterion.
\newline
The entropy inequality for \eqref{eq:iso_euler} reads in distributional form
\begin{align} \label{eq:entropy_inequality:iso}
 \partial_t E(\rho, m) + \partial_\spx \bigl( (E(\rho, m) + p(\rho)) v \bigr) \leq 0,
\end{align}
with the mathematical entropy function $E(\rho, m) \coloneqq \rho \, \mu(\rho) + \tfrac{m^2}{2\rho}$.
Here, $\mu(\rho) \coloneqq -RT_{\mathrm{ref}} \ln(\rho^{-1} - b) - a \rho$ denotes the corresponding specific Helmholtz free energy for the van der Waals pressure \eqref{eq:vdw_pressure}.
However, due to the (global) non-convexity of the pressure function \eqref{eq:vdw_pressure}, the entropy inequality \eqref{eq:entropy_inequality:iso} is insufficient for selecting unique entropy solutions --- see \cite{abeyaratne.knowles:evolution:2006,truskinovsky:kinks:1993,lefloch:hyperbolic:2002}.
Nevertheless, the well-posedness of the problem can be restored by employing another entropy criterion, such as Liu's entropy criterion \cite{liu:riemann:1974}.
Then, the Riemann problem \eqref{eq:riemannproblem} with $\vu_{\phasel} \in \cP_\phasel$ and $\vu_{\phaser} \in \cP_\phaser$ has a unique solution \cite{godlewski.seguin:riemann:2006}, which is consistent with the entropy inequality \eqref{eq:entropy_inequality:iso} --- see \cite{dafermos:hyperbolic:2016}.

To solve the Riemann problem \eqref{eq:riemannproblem} for \eqref{eq:iso_euler} with $\vu_{\phasel} \in \cP_\phasel$ and $\vu_{\phaser} \in \cP_\phaser$, we apply the Riemann solver described in \cite{rohde.zeiler:riemann:2018} (using kinetic relation $K_7$ in \cite{rohde.zeiler:riemann:2018} corresponding to Liu's entropy criterion and neglecting surface tension effects).
The Riemann solver yields the mapping
\begin{align} \label{eq:riemannsolver:iso_euler}
 \cR_{\mathrm{2p}} \colon \cP_{\phasel} \times \cP_{\phaser} \to  \bR^2 \times \bR^2 \times \bR:
 ((\rho_{\phasel}, m_{\phasel}), (\rho_{\phaser},m_{\phaser})) \mapsto ((\rho^{*}_{\phasel}, m^{*}_{\phasel}), (\rho^{*}_{\phaser}, m^{*}_{\phaser}), s),
\end{align}
by means of extracting the phase boundary wave states $\vu^{*}_{\phasel} \coloneqq (\rho^{*}_{\phasel}, m^{*}_{\phasel})$, $\vu^{*}_{\phaser} \coloneqq (\rho^{*}_{\phaser}, m^{*}_{\phaser})$ and speed $s$ from the Riemann solution for the Riemann initial condition \eqref{eq:riemannproblem} with $\vu_{\phasel} \coloneqq (\rho_{\phasel}, m_{\phasel})$ and $\vu_{\phaser} \coloneqq (\rho_{\phaser},m_{\phaser})$.

In this setting we seek to build a learned model that approximates $\cR_{\mathrm{2p}}$ while trying to fulfill the Rankine--Hugoniot conditions \eqref{eq:rh:iso}.
Thus, the constraint target function takes the form
\begin{align} \label{eq:constraint:iso}
 \Phi_{\mathrm{2p}}(\rho^{*}_{\phasel}, m^{*}_{\phasel}, \rho^{*}_{\phaser}, m^{*}_{\phaser}, s)
 &\coloneqq \abs[\big]{m^{*}_{\phasel} - m^{*}_{\phaser} - s (\rho^{*}_{\phasel} - \rho^{*}_{\phaser})}  \\
 &\quad + \abs[\big]{\tfrac{1}{\rho^{*}_{\phasel}} m^{*}_{\phasel} \cdot m^{*}_{\phasel} + p^{*}_{\phasel} - \tfrac{1}{\rho^{*}_{\phaser}} m^{*}_{\phaser} \cdot m^{*}_{\phaser} - p^{*}_{\phaser} - s (m^{*}_{\phasel} - m^{*}_{\phaser}) }.
\end{align}
Due to the highly nonlinear nature of the constraint \eqref{eq:constraint:iso}, we refrain from the analytic resolution of the constraint as described in \autoref{sec:constraint_layer_net}.
Instead, we construct an approximation of the function $\Psi$.
By multiplying the general Rankine--Hugoniot condition \eqref{eq:rh_condition} with $\vu^{*}_{\phasel} - \vu^{*}_{\phaser}$ we arrive at the least squares approximation for $s$ given $f(\vu^{*}_{\phasel}) - f(\vu^{*}_{\phaser})$ and $\vu^{*}_{\phasel} - \vu^{*}_{\phaser}$:
\begin{align} \label{eq:least_squares_speed}
 \widetilde{s} = \widetilde{s}(\vu^{*}_{\phasel}, \vu^{*}_{\phaser}) \coloneqq \frac{\bigl( f(\vu^{*}_{\phasel}) - f(\vu^{*}_{\phaser}) \bigr) \cdot (\vu^{*}_{\phasel} - \vu^{*}_{\phaser})}{ \norm{\vu^{*}_{\phasel} - \vu^{*}_{\phaser}}^2_2 },
\end{align}
for $\vu^{*}_{\phasel} \neq \vu^{*}_{\phaser}$.
We use formula \eqref{eq:least_squares_speed}
to define the approximate resolving function
\begin{align} \label{eq:psi:iso}
 \Psi_{\mathrm{2p}}(\vu^{*}_{\phasel}, \vu^{*}_{\phaser}) \coloneqq \widetilde{s}(\vu^{*}_{\phasel}, \vu^{*}_{\phaser}),
\end{align}
which will be used in the resolving layer method --- see \autoref{sec:constraint_layer_net}.
Clearly, \eqref{eq:least_squares_speed} is only a necessary condition for the constraint target function $\Phi_{\mathrm{2p}}$ to fulfill $\Phi_{\mathrm{2p}}(\rho^{*}_{\phasel}, m^{*}_{\phasel}, \rho^{*}_{\phaser}, m^{*}_{\phaser}, s) = 0$.

\subsection{Euler Equations: Contact Discontinuity}
\label{sec:euler_model}

Neglecting viscosity and heat conduction, a compressible fluid flow
is described in one spatial dimension by the Euler equations
\begin{align} \label{eq:euler}
\partial_t \begin{pmatrix} \rho \\ m \\ E\end{pmatrix}
+ \partial_\spx \begin{pmatrix} m \\ \tfrac{1}{\rho} m^2 + p(\rho, \varepsilon) \\ (E + p(\rho, \varepsilon)) \tfrac{m}{\rho} \end{pmatrix}
= \vec{0}
\quad \Leftrightarrow \quad
\partial_t \vu + \partial_\spx f(\vu) = \vec{0},
\end{align}
with fluid density $\rho$, momentum $m = \rho v$ and total energy density $E$ as the unknowns.
We will write $\vu = (\rho, m, E)$ for the vector of the conservative variables.
The total energy density can be written as $E = \rho \varepsilon + \tfrac{1}{2 \rho} \abs{m}^2$, where $\varepsilon$ is the specific internal energy of the fluid.
We consider an ideal gas, and the pressure function takes the form
\begin{align} \label{eq:ideal_gas}
 p(\rho, \varepsilon) \coloneqq (\gamma - 1) \rho \varepsilon,
\end{align}
with heat capacity ratio $\gamma = 1.4$.
In particular, by using \eqref{eq:ideal_gas} the system is hyperbolic and therefore a model problem for \eqref{eq:conservation_law}.

We consider the Cauchy problem for \eqref{eq:euler} with Riemann initial data \eqref{eq:riemannproblem}
with states $\vu_{\phaselr} \in \cP_{\phaselr} \subset \bR_+ \times \bR \times \bR_+$ that do not produce vacuum, i.e. those pairs that fulfill the pressure positivity condition
\begin{align} \label{eq:positivity_condition}
 \tfrac{2}{\gamma -1} \sqrt{\tfrac{\gamma p_\phasel}{\rho_\phasel}}
 + \tfrac{2}{\gamma -1} \sqrt{\tfrac{\gamma p_\phaser}{\rho_\phaser}}
 > v_\phaser - v_\phasel,
\end{align}
see \cite{toro:riemann:2009} for more details.

In this case, the Riemann solution always consists of three elementary waves one of which is always corresponding to a contact discontinuity --- see  \autoref{fig:wave_pattern:euler}.
The contact discontinuity is a discontinuous wave belonging to the second characteristic field of the flux function for \eqref{eq:euler}, which is linearly-degenerate.
As such, the entropy dissipation vanishes, and the pressure and fluid velocity stay constant across the jump.
In numerical simulations, contact discontinuities are especially prone to numerical diffusion and therefore difficult to resolve accurately.
Consequently, we focus on the contact discontinuity as the \emph{wave of interest} for this model problem.

To solve the Riemann problem for \eqref{eq:euler} we apply the Riemann solver described in \cite{toro:riemann:2009}.
Similar to the two previous models, we obtain a mapping
\begin{align} \label{eq:riemannsolver:euler}
 & \cR_{\mathrm{Euler}} \colon \cP_{\phasel} \times \cP_{\phaser} \to  \bR^3 \times \bR^3 \times \bR: \\
 &\quad \bigl((\rho_{\phasel}, m_{\phasel}, E_{\phasel}), (\rho_{\phaser}, m_{\phaser}, E_{\phaser}) \bigr) \mapsto \bigl((\rho^{*}_{\phasel}, m^{*}_{\phasel}, E^{*}_{\phasel}), (\rho^{*}_{\phaser}, m^{*}_{\phaser}, E^{*}_{\phaser}), s\bigr),
\end{align}
that provides the speed $s$ of and the states $\vu^{*}_{\phasel} \coloneqq (\rho^{*}_{\phasel}, m^{*}_{\phasel}, E^{*}_{\phasel})$, $\vu^{*}_{\phaser} \coloneqq (\rho^{*}_{\phaser}, m^{*}_{\phaser}, E^{*}_{\phaser})$ on either side of the contact discontinuity which  evolves from Riemann initial data $(\vu_{\phasel}, \vu_{\phaser}) \coloneqq ((\rho_{\phasel}, m_{\phasel}, E_{\phasel}), (\rho_{\phaser}, m_{\phaser}, E_{\phaser}))$.

The Rankine--Hugoniot jump conditions for \eqref{eq:euler} are
\begin{align} \label{eq:rh:euler}
\begin{aligned}
  \jump[\big]{ m - s \rho } &= 0, \\
  \jump[\big]{\tfrac{1}{\rho} m \cdot m + p(\rho, \varepsilon) - s m } &= 0, \\
  \jump[\big]{ (E + p(\rho, \varepsilon)) \tfrac{m}{\rho} - s E } &= 0.
\end{aligned}
\end{align}

Again, we want to find an constraint--aware approximate model for $\cR_{\mathrm{Euler}}$.
In this case, the constraint target function is
\begin{align} \label{eq:constraint:euler}
 \MoveEqLeft \Phi_{\mathrm{Euler}}(\rho^{*}_{\phasel}, m^{*}_{\phasel}, E^{*}_{\phasel}, \rho^{*}_{\phaser}, m^{*}_{\phaser}, E^{*}_{\phaser}, s)  \\
 &= \abs[\big]{m^{*}_{\phasel} - m^{*}_{\phaser} - s (\rho^{*}_{\phasel} - \rho^{*}_{\phaser})}  \\
 &\quad + \abs[\big]{\tfrac{1}{\rho^{*}_{\phasel}} m^{*}_{\phasel} \cdot m^{*}_{\phasel} + p^{*}_{\phasel} - \tfrac{1}{\rho^{*}_{\phaser}} m^{*}_{\phaser} \cdot m^{*}_{\phaser} - p^{*}_{\phaser} - s (m^{*}_{\phasel} - m^{*}_{\phaser}) } \\
 &\quad + \abs[\big]{ (E^{*}_{\phasel} + p^{*}_{\phasel}) \tfrac{m^{*}_{\phasel}}{\rho^{*}_{\phasel}} - (E^{*}_{\phaser} + p^{*}_{\phaser}) \tfrac{m^{*}_{\phaser}}{\rho^{*}_{\phaser}}  - s (E^{*}_{\phasel} - E^{*}_{\phaser}) },
\end{align}
which corresponds to the mass, momentum and energy conservation of \eqref{eq:rh:euler}.

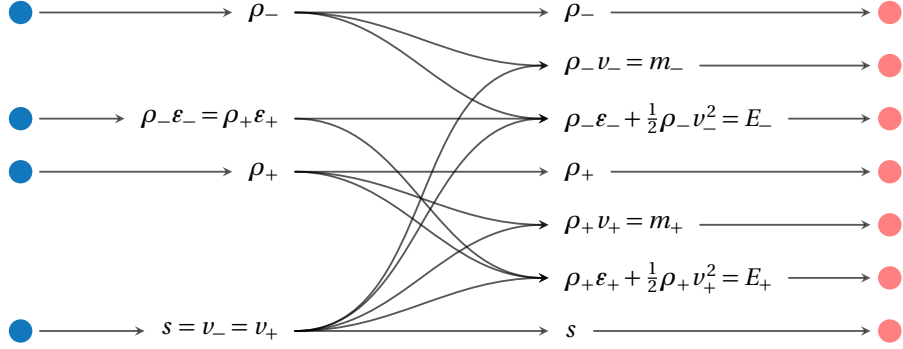
\begin{figure}[ht]
\centering
\resizebox{0.8\columnwidth}{!}{
\tikzsetnextfilename{resolving_layer_euler}
\begin{tikzpicture}[
font=\sffamily,
scale = 2.0,
node distance=8mm,
every node/.style = {on grid, anchor=center},
arrow/.style={->, opacity=0.66, >=stealth, thick, line join=round, line cap=round, shorten >=3pt, shorten <=3pt},
connection/.style={opacity=0.66, >=stealth, thick, line join=round, line cap=round, shorten >=3pt, shorten <=3pt}
]
\tikzstyle{neuron}=[circle,fill=black!25,minimum size=10pt,inner sep=0pt]
\tikzstyle{input neuron}=[neuron, fill=NavyBlue!90];
\tikzstyle{output neuron}=[neuron, fill=red!50];
\tikzstyle{left column}=[align=right, anchor=east];
\tikzstyle{right column}=[align=left, anchor=west];

\node[input neuron] (a0) at (-2.0,0) {};
\node[below of=a0] (a1) {};
\node[below of=a1, input neuron] (a2) {};
\node[below of=a2, input neuron] (a3) {};
\node[below of=a3] (a4) {};
\node[below of=a4] (a5) {};
\node[below of=a5, input neuron] (a6) {};

\node[left column] (in0) at (0,0) {$\rho_\phasel$};
\node[below=of in0.east, left column] (in1) {};
\node[below=of in1.east, left column] (in2) {$\rho_\phasel\varepsilon_\phasel = \rho_\phaser\varepsilon_\phaser$};
\node[below=of in2.east, left column] (in3) {$\rho_\phaser$};
\node[below=of in3.east, left column] (in4) {};
\node[below=of in4.east, left column] (in5) {};
\node[below=of in5.east, left column] (in6) {$s = v_\phasel = v_\phaser$};

\draw[arrow] (a0.east) to[out=0,in=180] (in0.west);
\draw[arrow] (a2.east) to[out=0,in=180] (in2.west);
\draw[arrow] (a3.east) to[out=0,in=180] (in3.west);
\draw[arrow] (a6.east) to[out=0,in=180] (in6.west);

\node[right column] (out0) at (2.0,0) {$\rho_\phasel$};
\node[below=of out0.west, right column] (out1) {$\rho_\phasel v_\phasel = m_\phasel$};
\node[below=of out1.west, right column] (out2) {$\rho_\phasel \varepsilon_\phasel + \tfrac{1}{2} \rho_\phasel v^2_\phasel = E_\phasel$};
\node[below=of out2.west, right column] (out3) {$\rho_\phaser$};
\node[below=of out3.west, right column] (out4) {$\rho_\phaser v_\phaser = m_\phaser$};
\node[below=of out4.west, right column] (out5) {$\rho_\phaser \varepsilon_\phaser + \tfrac{1}{2} \rho_\phaser v^2_\phaser = E_\phaser$};
\node[below=of out5.west, right column] (out6) {$s$};

\draw[arrow] (in0.east) to[out=0,in=180] (out0.west);
\draw[arrow] (in0.east) to[out=0,in=180] (out1.west);
\draw[arrow] (in0.east) to[out=0,in=180] (out2.west);
\draw[arrow] (in2.east) to[out=0,in=180] (out2.west);
\draw[arrow] (in2.east) to[out=0,in=180] (out5.west);
\draw[arrow] (in3.east) to[out=0,in=180] (out3.west);
\draw[arrow] (in3.east) to[out=0,in=180] (out4.west);
\draw[arrow] (in3.east) to[out=0,in=180] (out5.west);
\draw[arrow] (in6.east) to[out=0,in=180] (out6.west);
\draw[arrow] (in6.east) to[out=0,in=180] (out1.west);
\draw[arrow] (in6.east) to[out=0,in=180] (out2.west);
\draw[arrow] (in6.east) to[out=0,in=180] (out4.west);
\draw[arrow] (in6.east) to[out=0,in=180] (out5.west);

\node[output neuron] (b0) at (4.5,0) {};
\node[below of=b0, output neuron] (b1) {};
\node[below of=b1, output neuron] (b2) {};
\node[below of=b2, output neuron] (b3) {};
\node[below of=b3, output neuron] (b4) {};
\node[below of=b4, output neuron] (b5) {};
\node[below of=b5, output neuron] (b6) {};

\draw[arrow] (out0.east) to[out=0,in=180] (b0.west);
\draw[arrow] (out1.east) to[out=0,in=180] (b1.west);
\draw[arrow] (out2.east) to[out=0,in=180] (b2.west);
\draw[arrow] (out3.east) to[out=0,in=180] (b3.west);
\draw[arrow] (out4.east) to[out=0,in=180] (b4.west);
\draw[arrow] (out5.east) to[out=0,in=180] (b5.west);
\draw[arrow] (out6.east) to[out=0,in=180] (b6.west);

\end{tikzpicture}
}
\caption{Sketch of the constraint resolving layer for the Euler model.}
\label{fig:resolving:euler}
\end{figure}

To construct the resolving layer of the \cres-method, we exploit the fact that the pressure $p$ and the velocity $v$ are constant across the contact discontinuity.
Then, it follows from the Rankine--Hugoniot conditions \eqref{eq:rh:euler} that
$s = v_\phasel = v_\phaser$ and $\rho_\phasel \varepsilon_\phasel = \rho_\phaser \varepsilon_\phaser$ holds.
Consequently, we can reconstruct the whole set of variables from the quantities $\rho_\phasel$, $\rho_\phaser$, $(\rho_\phasel\varepsilon_\phasel)$ and $s$.
Motivated by the properties of the wave of interest, we can define the resolving function
\begin{align} \label{eq:psi:euler}
 \Psi_{\mathrm{Euler}}(\rho_\phasel, \rho_\phasel \varepsilon_{\phasel}, \rho_\phaser, s)
 &\coloneqq
 \bigl(\rho_\phasel,~
 \rho_\phasel s,~
 \rho_\phasel\varepsilon_\phasel + \tfrac{1}{2} \rho_\phasel s^2,~
 \rho_\phaser,~
 \rho_\phasel s,~
 \rho_\phasel\varepsilon_\phasel + \tfrac{1}{2} \rho_\phaser s^2,~
 s \bigr) \\
 &= (\rho_{\phasel}, m_{\phasel}, E_{\phasel}, \rho_{\phaser}, m_{\phaser}, E_{\phaser}, s).
\end{align}
\autoref{fig:resolving:euler} contains a graphical representation of the resolving layer.

\section{Application of Constraint-Aware Learning Methods to Case Study Model Problems}
\label{sec:technical_details}

In this section we provide details on how the constraint-aware learning methods are applied to the case study model problems of \autoref{sec:models} and the numerical simulations.
First we describe the data set generation, then the network training, and finally the numerical discretization scheme.

\subsection{Data Generation}
\label{sec:data_generation}

To generate training data sets $D_{\mathrm{train}} = \{ (\ix_i, \oy_i) : i=1, \ldots, N_{\mathrm{train}}\}$ with $N_{\mathrm{train}}$ samples, we evaluate the Riemann solver \eqref{eq:riemannsolver:generic} on input values $X_{\mathrm{train}} = \{\ix_i : i = 1, \ldots, N_{\mathrm{train}}\}$ with $\ix_i = (\vu_{\phasel, i}, \vu_{\phaser, i}) \in \cX$.
The set $X_{\mathrm{train}}$ is generated in a random way for a given bounded domain $B \subset \cX$ using a variation of Mitchell's algorithm \cite{mitchell:spectrally:1991} to generate evenly and randomly distributed points in $\bR^{d_{\mathrm{in}}}$.
Evaluation of the input points $\ix_i \in X_{\mathrm{train}}$ by the Riemann solver \eqref{eq:riemannsolver:generic} yields the output value set $Y_{\mathrm{train}} = \{\oy_i : i = 1, \ldots, N_{\mathrm{train}}\}$ with $\oy_i = (\vu^{*}_{\phasel,i}, \vu^{*}_{\phaser,i}, s_i) = \cR(\vu_{\phasel,i}, \vu_{\phaser,i})$, for $i=1, \ldots, N_{\mathrm{train}}$.

Furthermore, we seek to generate noisy data sets, to simulate Riemann solutions from approximate models, such as particle simulations.
To generate noisy data, we start with computing the empirical standard deviation $\overline{\vec{\sigma}}_j$ over $Y_{\mathrm{train}}$ for each component $j = 1,\ldots,d_{\mathrm{out}}$ of the output values $\oy \in Y_{\mathrm{train}}$.
Then we introduce a noise level parameter $l_\mathrm{noise} \geq 0$ and transform the output values $\oy_i \in Y_{\mathrm{train}}$ as
\begin{align}
 \oy_{i,j} \mapsto  \oy_{i,j} + l_\mathrm{noise} \, \overline{\vec{\sigma}}_j \, R,
\end{align}
where $R$ is an uniformly distributed random variable on $[-1,1]$.
In this way we can generate training data sets $D_{\mathrm{train}}$ with an arbitrary noise level $l_\mathrm{noise}$.
\newline
For the test data sets $D_{\mathrm{test}}$, the input data set $X_{\mathrm{test}}$ consists of a uniform grid over the bounded domain $B \subset \cX$ on which the Riemann solver is evaluated.

\subsection{Technical Details of the Network Training}
\label{sec:training_details}

In this section we describe the technical details of the neural network training.
\newline
All training was performed withing the neural network framework PyTorch \cite{paszke.gross.ea:automatic:2017}.
A tabular overview of the parameters used for the results presented in \autoref{sec:results} is given in \autoref{table:net_parameters}.

Before training each network, we standardize the input samples of each data set by subtracting the mean and dividing by the standard deviation.
Then, we randomly split 20\% from the initial training data set to form a validation data set $D_{\mathrm{val}}$.
During the training process, we evaluate the loss on the validation data set and we will save the network with the best validation loss (over all epochs) to avoid overfitting.
As mentioned in \autoref{sec:neural_networks}, we apply $\ell^2$-weight decay.

For the optimization procedure we use the Adam optimizer \cite{kingma.ba:adam:2014} with a prescribed learning rate.
The batch size is chosen  for each data set as the largest power of 2 such that we have at least \num{5} batches.
The weights of the networks are initialized by applying the Glorot normal initializer \cite{glorot.bengio:understanding:2010}, the bias parameters are initially set to zero.

During the training procedure not-a-number values might be encountered, as for example due to the arbitrariness of the network output values a division by values close to zero might occur inside the constraint target function of \eqref{eq:constraint_adapted_loss}.
In case any not-a-number values are encountered during training, we stop the current epoch, decrease the learning rate and resume training from the last valid snapshot, with the lowest validation loss that was encountered up to this point.
To further stabilize the scheme, we apply gradient norm clipping \cite{pascanu.mikolov.ea:difficulty:2013} with a clip value of 1.

\begin{table}[h]
\sisetup{
    table-text-alignment=center,
    table-number-alignment=center,
    table-figures-exponent=1,
    table-align-exponent=false,
}
\centering
\begin{tabular}{p{9em}*{3}{S[table-format=1e-1]}}
\toprule
& \multicolumn{3}{c}{\bfseries Model problems} \\ \cmidrule{2-4}
\bfseries Parameter &
\multicolumn{1}{>{\centering}m{6em}}{\bfseries Cubic} &
\multicolumn{1}{>{\centering}m{8em}}{\bfseries Two-phase flow} &
\multicolumn{1}{>{\centering}m{6em}}{\bfseries Euler}\\ \midrule
hidden layers & 5 & 5 & 7 \\
nodes per hidden layer & 20 & 30 & 70 \\
learning rate & 5e-4 & 2e-4 & 5e-5 \\
max epochs & 5e4 & 5e4 & 2e6 \\
patience $N_{\mathrm{patience}}$ & 5e3 & 5e3 & e4 \\
weight decay $\alpha_{\mathrm{wd}}$ & e-7 & e-7 & e-9 \\
\bottomrule
\end{tabular}%
\caption{List of parameters used in \autoref{sec:results}.}
\label{table:net_parameters}
\end{table}

\subsection{Finite Volume Moving Mesh Front Capturing Scheme}
\label{sec:front_capturing_scheme}

\begin{figure}[ht]
 \centering
\resizebox{0.66\columnwidth}{!}{
\tikzsetnextfilename{moving_mesh_1d}
\begin{tikzpicture}[scale = 2.5]

\tikzset{facestyle/.style={fill=lightgray, draw=black, very thin, line join=round, fill opacity=0.1, draw opacity=1.0}}
\tikzset{facestylebackside/.style={shade,  fill=lightgray, very thin, line join=round, fill opacity=0.33, draw opacity=0}}

\tikzset{coordinateaxis/.style={->, >=stealth, thick, line join=round, line cap=round}}
\tikzset{tickline/.style={-, line join=round, line cap=round}}
\tikzset{rarefactionline/.style={-, line join=round, line cap=round, opacity=0.5}}

\fill[NavyBlue!10] (-1.5, 0) -- (-0.1, 0.0) -- (0.15, 0.5) -- (-1.5, 0.5) -- cycle;

\draw[coordinateaxis] (-1.7,0) -- (-1.7,0.5) node[pos=1.0, above] {$t^{n+1}$} node[pos=0.0, below] {$t^{n}$};

\draw[coordinateaxis] (-1.5,0.5) -- (1.5,0.5) node[pos=1.0, right] {};
\draw[coordinateaxis] (-1.5,0) -- (1.5,0) node[pos=1.0, right] {};

\draw[tickline] (-1.0, -0.025) -- (-1.0, 0.025) node[pos=0.0, below]{$\spx_{i-1}$};
\draw[tickline] (-0.5, -0.025) -- (-0.5, 0.025) node[pos=0.0, below]{$\spx_{i}$};

\draw[NavyBlue, very thick, tickline] (-0.1, -0.025) -- (-0.1, 0.025) node[pos=0.0, below]{$\spx_{\Gamma}^{n}$};

\draw[tickline] (0.5, -0.025) -- (0.5, 0.025) node[pos=0.0, below]{$\spx_{i+1}$};
\draw[tickline] (1.0, -0.025) -- (1.0, 0.025) node[pos=0.0, below]{$\spx_{i+2}$};

\draw[tickline] (-1.0, {0.5-0.025}) -- (-1.0, {0.5+0.025});
\draw[tickline] (-0.5, {0.5-0.025}) -- (-0.5, {0.5+0.025});

\draw[NavyBlue, very thick, tickline] (0.15, {0.5-0.025}) -- (0.15, {0.5+0.025}) node[pos=0.0, above]{$\spx_{\Gamma}^{n+1}$};

\draw[tickline] (0.5, {0.5-0.025}) -- (0.5, {0.5+0.025});
\draw[tickline] (1.0, {0.5-0.025}) -- (1.0, {0.5+0.025});

\draw[dotted, line join=round, line cap=round, opacity=0.5] (-1.0, 0.05) -- (-1.0, 0.45);
\draw[dotted, line join=round, line cap=round, opacity=0.5] (-0.5, 0.05) -- (-0.5, 0.45);
\draw[dotted, line join=round, line cap=round, opacity=0.5] (0.5, 0.05) -- (0.5, 0.45);
\draw[dotted, line join=round, line cap=round, opacity=0.5] (1.0, 0.05) -- (1.0, 0.45);

\draw[NavyBlue, ->, >=stealth, thick, line join=round, line cap=round, shorten >=1ex, shorten <=1ex] (-0.1, 0) -- (0.15, 0.5) node[pos=0.5, right]{$s$};

\node[NavyBlue, anchor=center] at (-0.75,0.25) {$\cP_\phasel$};
\node[NavyBlue, anchor=center] at (0.75,0.25) {$\cP_\phaser$};

\end{tikzpicture}
}
\caption{Sketch of the one-dimensional front capturing scheme following a discontinuous wave connecting the $\cP_\phasel$-domain and the $\cP_\phaser$-domain.}
\label{fig:1d_front_capturing}
\end{figure}
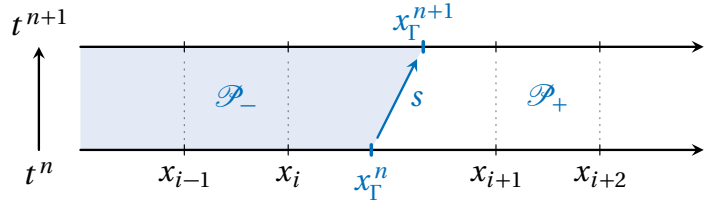

To implement the numerical discretization of conservation laws while resolving discontinuous wave fronts within the mesh, we apply the one-dimensional version of the front-capturing scheme described in \cite{chalons.rohde.ea:finite:2017}.
It is a finite volume scheme with moving edges, such that the position $\spx_\Gamma(t)$ of a discontinuous wave is always be fully resolved within the mesh.
In our case, we seek to keep track of discontinuous waves that connect regions with values in $\cP_\phasel$ and regions with values in $\cP_\phaser$ as illustrated in \autoref{fig:1d_front_capturing}.
In the following we present the most relevant parts of the algorithm --- for more details we refer to \cite{chalons.rohde.ea:finite:2017}.

The finite volume method on a time-dependent, one-dimensional mesh
for the system \eqref{eq:conservation_law} is of the form
\begin{align} \label{eq:fv-scheme}
 \vu^{n+1}_{i} = \vu^{n}_{i} - \frac{\Delta t}{\Delta \spx^n_i} \bigl( \vec{g}^{n}_{i + \sfrac{1}{2}} - \vec{g}^{n}_{i - \sfrac{1}{2}} \bigr).
\end{align}
Here, $\vu^{n}_{i}$ denotes the $i$th cell average of the unknown $\vu(\spx,t)$ at the time $t^n \geq 0$.
The time step is given by $\Delta t$, the $i$th cell width at time $t^n$ is denoted by $\Delta \spx^n_i > 0$.
The terms $g^{n}_{i \pm \sfrac{1}{2}}$ denote the numerical fluxes. 
If, at the $i$th cell, the wave of interest lies on the right-hand side, i.e. $\vu^n_{i} \in \cP_\phasel$ and $\vu^n_{i+1} \in \cP_\phaser$, the right numerical flux is given by
\begin{align} \label{eq:interface_flux:right}
 \vec{g}^{n}_{i + \sfrac{1}{2}} =
 g_\phasel(\vu^n_{i}, \vu^n_{i+1}) &= f(\vu^{*}_{\phasel}) - s \cdot \vu^{*}_{\phasel}, ~
 \text{ with } ~ \cR(\vu^n_{i}, \vu^n_{i+1}) = (\vu^{*}_{\phasel}, \vu^{*}_{\phaser}, s).
\end{align}
Similarly, if the wave of interest lies on the left-hand side, i.e. $\vu^n_{i-1} \in \cP_\phasel$ and $\vu^n_{i} \in \cP_\phaser$, the left numerical flux is
\begin{align} \label{eq:interface_flux:left}
 \vec{g}^{n}_{i - \sfrac{1}{2}} =
 g_\phaser(\vu^n_{i-1}, \vu^n_{i}) &= f(\vu^{*}_{\phaser}) - s \cdot \vu^{*}_{\phaser}, ~
 \text{ with } ~ \cR(\vu^n_{i-1}, \vu^n_{i}) = (\vu^{*}_{\phasel}, \vu^{*}_{\phaser}, s).
\end{align}
For all other cell boundaries we apply the Lax--Friedrichs numerical flux
\begin{align} \label{eq:lax-friedrichs}
 \vec{g}^{n}_{i + \sfrac{1}{2}}
 = g_\mathrm{LF}\bigl( \vu^n_{i}, \vu^n_{i+1} \bigr)
 \coloneqq \tfrac{1}{2} \, \bigl(f\bigl(\vu^n_{i}\bigr) + f\bigl(\vu^n_{i+1}\bigr)\bigr) - \frac{\alpha_{\mathrm{LF}}}{2} \, \bigl(\vu^n_{i+1} - \vu^n_{i}\bigr),
\end{align}
with the Lax--Friedrichs parameter $\alpha_{\mathrm{LF}} > 0$.

Note, that we split the numerical flux at the interface in two parts, one \eqref{eq:interface_flux:left} for the $\cP_\phasel$-domain and another one \eqref{eq:interface_flux:right} for the $\cP_\phaser$-domain.
Nevertheless, it holds that  $g_\phasel(\vu_{\phasel}, \vu_{\phaser}) = g_\phaser(\vu_{\phasel}, \vu_{\phaser})$ if the Rankine--Hugoniot condition \eqref{eq:rh_condition} is fulfilled.
\newline
To compute the numerical fluxes \eqref{eq:interface_flux:right}, \eqref{eq:interface_flux:left} at the interface, we need to evaluate the mapping $\cR$, i.e. solve the Riemann problem for the initial data $(\vu_{\phasel}, \vu_{\phaser})$ and obtain $(\vu^{*}_{\phasel},\vu^{*}_{\phaser}, s)$.
This can be expensive for more complex models, and forms our actual motivation to learn a model that approximates the mapping $\cR$.
Furthermore, the wave speed $s$ describes the movement of the interface edge $\spx_\Gamma(t)$ that occurs after each time step.

\section{Numerical Evaluation of the Constraint-Aware Learning Methods}
\label{sec:results}

In this section we test the performance of the constraint-aware learning methods (\autoref{sec:constraint-aware-methods}).
We split the section in three parts, one for each of the model problems presented in \autoref{sec:models}.
In each part, we first study the performance of the constraint-aware neural networks generated with the \cres- and \cadl-method.
This includes testing the methods for differently-sized data sets and with regards to the influence of noisy data on the resulting networks.
Thereafter, we use the resulting networks for some simulations using the front capturing scheme from \autoref{sec:front_capturing_scheme} and compare the results with analytic Riemann solutions.

\subsection{Performance Evaluation Procedure}

To test the performance of the constraint-aware learning methods, we split the evaluation procedure in two parts.
First, we train the networks and test them on an unknown data set.
Secondly, we check the performance of the networks in simulations.

To train the networks, we generate the training data sets $D_{\mathrm{train}} \subset \cX \times \cY$ with different number of samples $N_\mathrm{train} = \abs{D} \in \bN$ and noise levels $l_\mathrm{noise} \geq 0$.
For each data set we train three types of networks
\begin{itemize}
\item standard MLP ignoring constraint-preserving mechanisms,
\item MLP with a constraint-resolving layer (see \autoref{sec:constraint_layer_net}),
\item MLP trained with a constraint-adapted loss for different values for the constraint penalty parameter $\lambda$ (see \autoref{sec:constraint_loss_net}).
\end{itemize}
As the training process is not deterministic, we train several instances for every type of network and test all of them. 
After training, we evaluate the mean-squared error and constraint deviation on a separate test data set $D_{\mathrm{test}} \subset \cX \times \cY$.
This will be the final benchmark for the networks, as the test data is unknown and was never seen during the training process.

In the second part of the evaluation, the neural networks are used in fluid dynamic simulations using the front capturing scheme from \autoref{sec:front_capturing_scheme}.
To be more specific, the neural networks are used to evaluate the numerical fluxes \eqref{eq:interface_flux:right}, \eqref{eq:interface_flux:left} at the wave front.
To test the different methods in this setting, we run simulations for Riemann initial conditions and compute the error with respect to the analytic Riemann solution.
To avoid any intrinsic bias towards a particular choice of initial data, we perturb the initial Riemann data and compute the mean error over several simulation runs.

\subsection{Cubic Flux Model}
\label{sec:cubic_results}

\begin{figure}[H]
\centering
\makebox[\textwidth][c]{
\includegraphics[width=1.1\columnwidth]{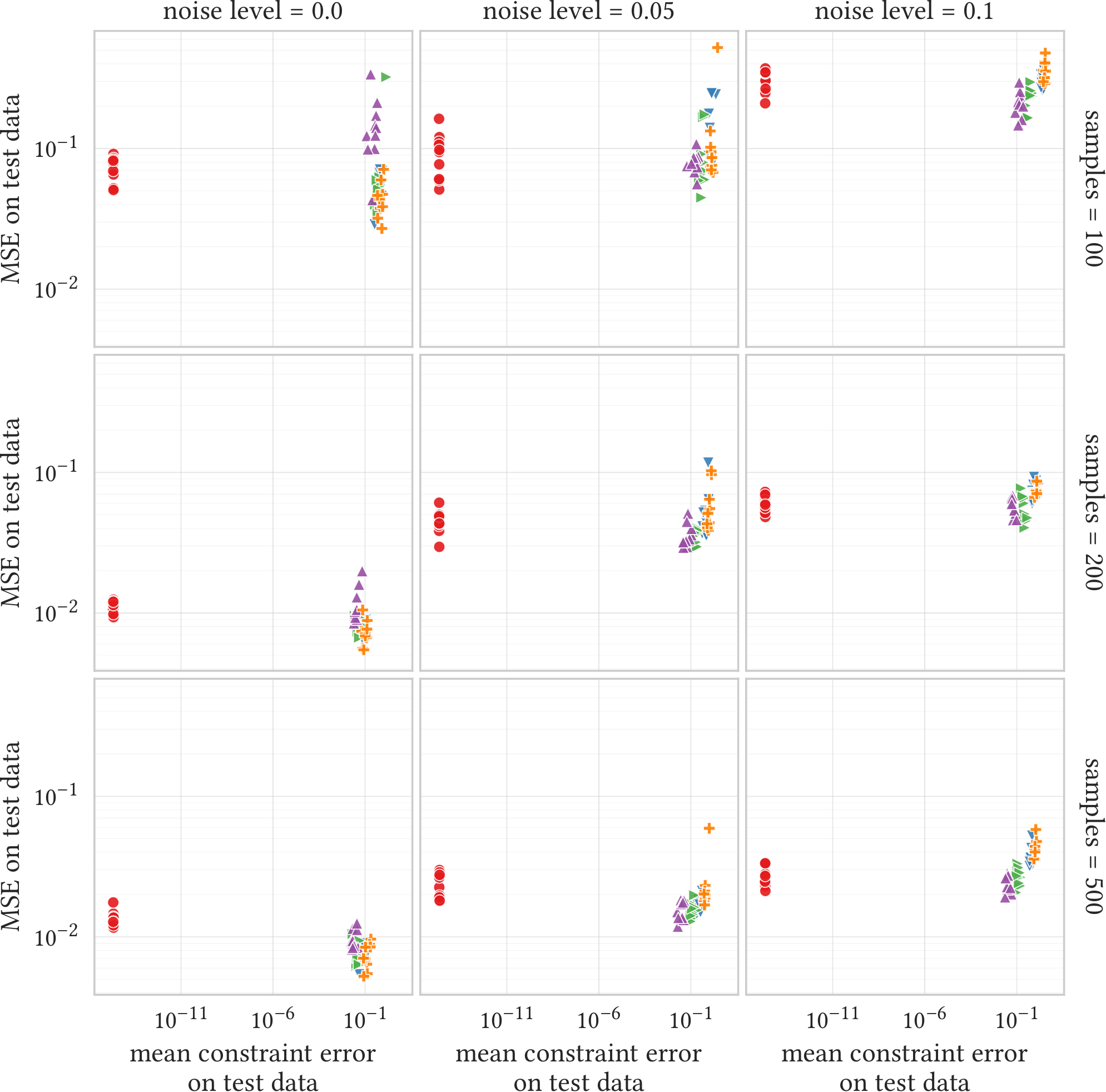}
}
\mbox{}

\includegraphics[width=0.9\columnwidth]{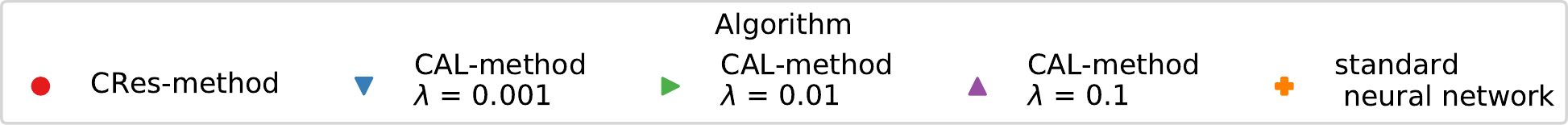}
\caption{Mean-squared errors and constraint errors on the test data set for the cubic flux model problem.
Each subfigure corresponds to one data set. \\
Horizontally the mean squared error on the test data set is plotted, vertically the mean constraint error.
Therefore, points that are located to the left and near bottom are considered better.
}
\label{fig:nn_benchmark:cubic}
\end{figure}

In this section we test the constraint-aware methods described in \autoref{sec:constraint-aware-methods} for the cubic flux model, i.e. we seek to learn the Riemann solver \eqref{eq:riemannsolver:cubic}, with $\kappa = \num{0.75}$ in the kinetic relation \eqref{eq:kinrel}.
The resulting networks are compared with their standard/constraint-unaware counterparts.
\newline
To this end, we consider data sets comprising $N_{\mathrm{train}} \in \{\num{100}, \num{200}, \num{500}\}$ samples, with noise levels $l_{\mathrm{noise}}\in \{\num{0}, \num{0.05}, \num{0.1}\}$.
These nine data sets are sampled as described in \autoref{sec:data_generation} on the bounded domain $B = [0, 5] \times [-2.5, 0]$.
The test data set $D_{\mathrm{test}}$ consists of \num{10000} samples in $B$.
On each data set we train a standard neural network, one with a constraint-resolving layer and a network with the constraint-adapted loss for each constraint penalty parameter value $\lambda \in \{\num{e-1}$, $\num{e-2}, \num{e-3}\}$.
As the training procedure is non-deterministic we train 10 separate instances for every type of network.
\newline
The mean-squared errors and mean constraint errors (in the $\ell^1$ metric) on the test data set are displayed in \autoref{fig:nn_benchmark:cubic}.
Concerning the constraint error, we can clearly see that the \cres-method outperforms the standard networks and the \cadl-method.
In case of noisy data sets, the \cadl-method seems to be able to produce (depending on the constraint penalty parameter $\lambda$) better results in terms of mean-squared errors and constraint errors than the standard approach.
The marginal improvement in the constraint error with the \cadl-method in the absence of noise, comes at the cost of a higher mean-squared error than the standard approach.

If we look at the mean-squared error, we see a difference between noisy and noise-free data sets as well.
If there is no noise, the standard networks and the \cadl-method behaves relatively similar
Only the \cres-method has a higher mean-squared error.
On the other side, if there is noise (especially for the high noise level \num{0.1}) the constraint aware methods yield a better mean-squared error  and constraint errors.

\begin{figure}[H]
 \centering
\resizebox{0.5\columnwidth}{!}{
\tikzset{external/export next=false}
\tikzsetnextfilename{riemannsolution_cubic}
\begin{tikzpicture}[scale=1.0, font=\sffamily]
    \begin{axis}[width=0.5\columnwidth, height=0.3\columnwidth,
    scale only axis,
    ymin=-1.2, ymax=1.2,  xlabel=$\frac{\spx}{t}$, ylabel=$u$,
    grid=both, minor x tick num=1, minor y tick num=1,
    major grid style={very thin, opacity=0.66},
    minor grid style={ultra thin, opacity=0.33},
    xmin=-1, xmax=4, samples=200]
    \addplot[NavyBlue, dashed, opacity=0.9] coordinates {
    (-5,1) (0,1) (0,-1) (5,-1)
    };
    \addplot[NavyBlue, very thick] table [x=s,y=u, col sep=comma] {data/cubic_riemann_solution.csv};
    \end{axis}
\end{tikzpicture}%
}
\caption{Reference Riemann solution for the cubic flux model problem.
The dashed line represents the initial conditions.}
\label{fig:riemann_solution:cubic}
\end{figure}
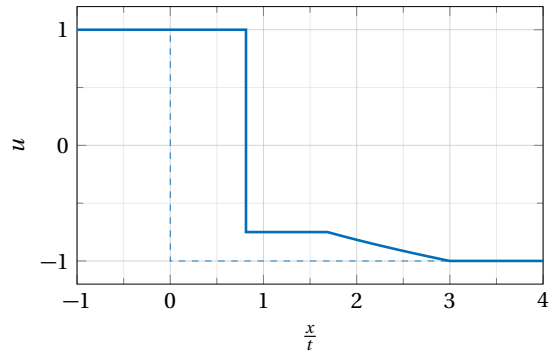

\begin{figure}[H]
\centering
\makebox[\textwidth][c]{
\includegraphics[width=1.05\columnwidth]{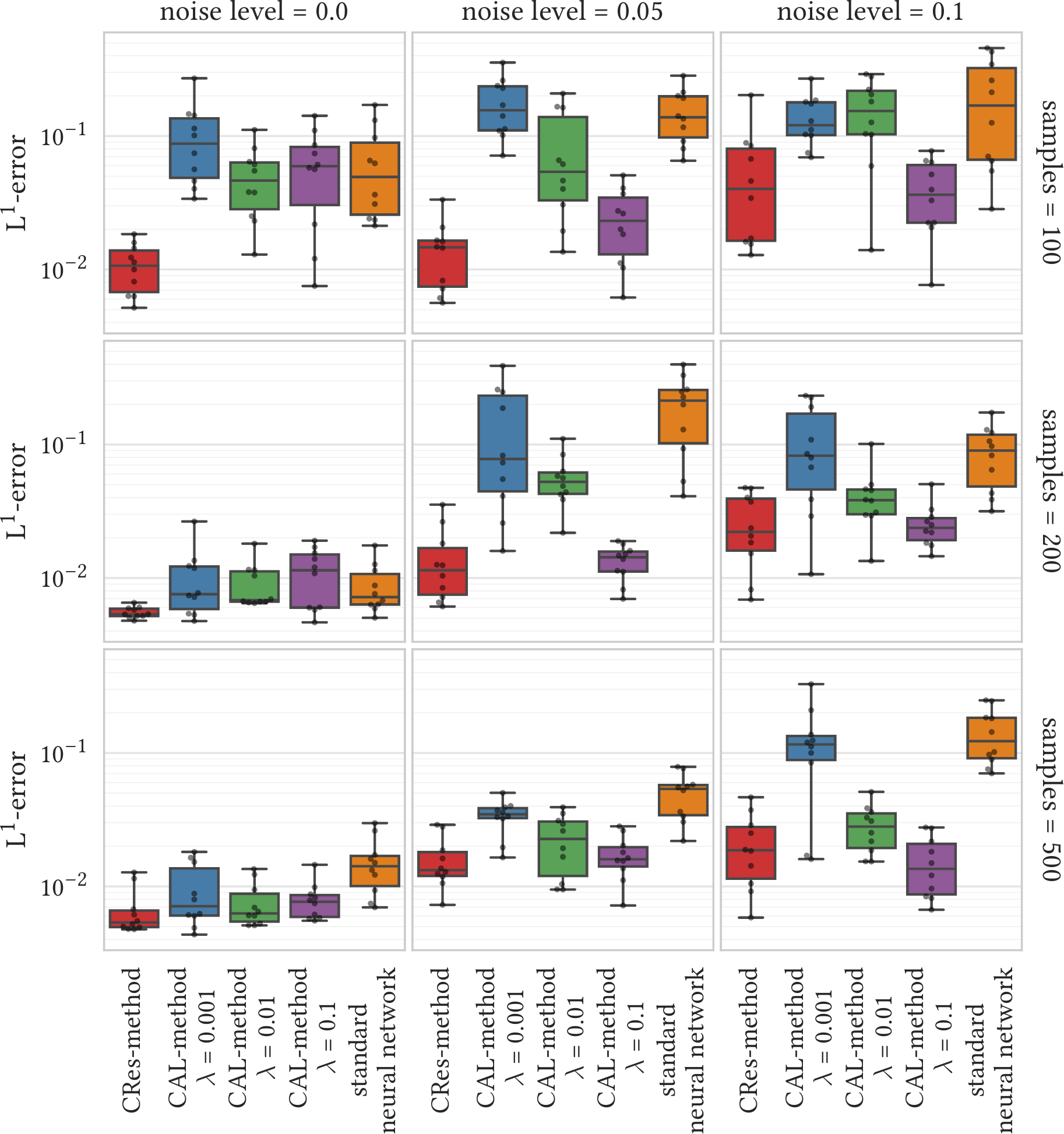}
}
\caption{Mean $L^1$-errors for the cubic flux model problem.
The mean $L^1$-errors of the finite volume simulation are computed over \num{25} perturbed Riemann problems for each trained network.
The gray dots represent the error for each network instance.
}
\label{fig:continuum_error:cubic}
\end{figure}

To test the neural networks in numerical simulations, we consider the Riemann problem \eqref{eq:riemannproblem} with $u_{\phasel} = -1$,  $u_{\phaser} = 1$.
The exact solution can be computed via the Riemann solver from \autoref{sec:cubic} and is displayed in \autoref{fig:riemann_solution:cubic}.
\newline
To avoid biased results due to the specific choice of $u_{\phasel}$ and $u_{\phaser}$,
we perturb the initial data with perturbations sampled uniformly from $(-0.1,0.1)$ and compute the mean of the $L^1$-errors over \num{25} runs.
The simulations are performed on a grid over the interval $(-2,2)$ with \num{2000} cells.
The timestep is $\Delta t = \num{2e-4}$ and we run the simulation up to $t_{\mathrm{end}} = \num{0.5}$.
For the numerical flux away from the interface we choose the Lax--Friedrichs flux with $\alpha_{\mathrm{LF}} = 2$.
For these parameters, if we apply the exact Riemann solver for the computation of the interface fluxes \eqref{eq:interface_flux:right} and \eqref{eq:interface_flux:left}, the absolute $L^1$-error amounts to approximately \num{4.8e-3}, which is the baseline  error of the discretization. %
\newline
We repeat the simulation for each network instance we have trained in the first part of the test and compute the $L^1$-error with respect to the exact Riemann solution.
The resulting $L^1$-errors are displayed in \autoref{fig:continuum_error:cubic}.

For all data sets, the \cres-method yields a lower mean error than the standard networks.
In case of noisy data sets, depending on the choice of the constraint penalty parameter $\lambda$, the \cadl-method is also able to produce much better results than the standard networks.
The standard networks work quite well on the noise-free data sets, therefore the gain by applying the \cadl-method is less prominent.

\pagebreak
\subsection{Isothermal Two-Phase Flow}
\label{sec:iso_results}

In this section we re-run the computational tests from the previous section for the isothermal two-phase flow model (\autoref{sec:iso_euler}).
\newline
The data sets have $N_{\mathrm{train}} \in \{\num{1000}, \num{2000}, \num{5000} \}$ samples, each with noise levels $l_{\mathrm{noise}} = \{\num{0}, \num{0.05}, \num{0.1} \}$.
The samples are generated for $\rho_\phasel \in (\num{1.7}, \num{2.6})$,  $v_{\phasel} \in (-1,1)$, $\rho_\phaser \in (\num{0.01}, \num{0.33})$,  $v_{\phaser} \in (-1,1)$ and then transformed to the fluid momentum $m_{\phaselr} = \rho_{\phaselr} v_{\phaselr}$.
For the \cadl-method, we consider the constraint penalty parameter values $\lambda \in \{\num{e-1}$, $\num{e-2}, \num{e-3}\}$.

\begin{figure}[H]
\centering
\makebox[\textwidth][c]{
\includegraphics[width=1.1\columnwidth]{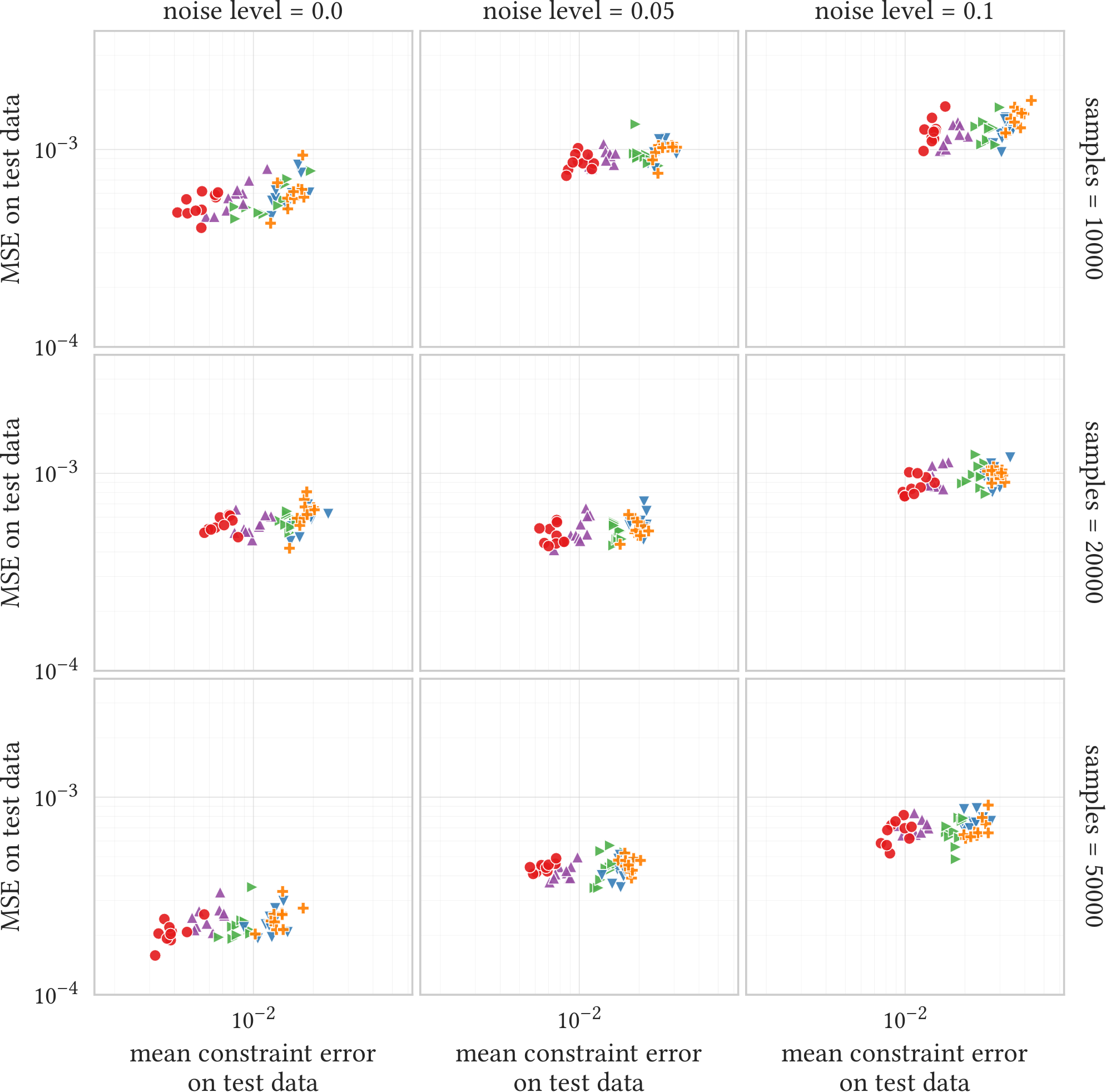}
}
\mbox{}

\includegraphics[width=0.9\columnwidth]{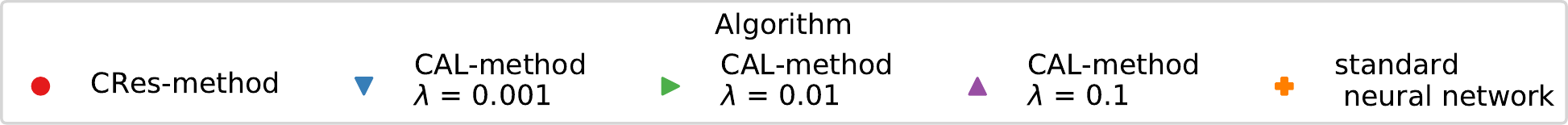}
\caption{Mean-squared errors and constraint errors on the test data set for the two-phase flow model problem.
Each subfigure corresponds to one data set. \\
Horizontally the mean squared error on the test data set is plotted, vertically the mean constraint error.
Therefore, points that are located to the left and near bottom are considered better.
}
\label{fig:nn_benchmark:iso}
\end{figure}

Again, we test standard networks and networks built with the \cadl- and the \cres-method on a test data set.
The results are presented in \autoref{fig:nn_benchmark:iso}.
Concerning the mean-squared error on the test data set, the standard networks, the \cadl- and the \cres-method perform similarly.
However, in terms of the constraint error on the test data set,
we observe that the \cres-method performs the best, despite using an approximate resolving function \eqref{eq:psi:iso}.
The \cadl-method works second best; for $\lambda = \num{e-1}$ it approaches the performance of the \cres-method, and for $\lambda = \num{e-3}$ it behaves almost like standard networks.

\begin{figure}[H]
 \centering
\resizebox{0.5\columnwidth}{!}{
\tikzset{external/export next=false}
\tikzsetnextfilename{riemannsolution_iso}
\begin{tikzpicture}[scale=1.0, font=\sffamily]
\begin{groupplot}[
    group style={
        group name=my plots,
        group size=1 by 2,
        x descriptions at=edge bottom,
        y descriptions at=edge left,
        horizontal sep=0.5cm,
        vertical sep=0.5cm,
    },
    width=0.5\columnwidth, height=0.3\columnwidth,
    scale only axis,
    grid=both, minor x tick num=1, minor y tick num=1,
    major grid style={very thin, opacity=0.66},
    minor grid style={ultra thin, opacity=0.33},
    xmin=-3, xmax=2,
    xlabel=$\frac{\spx}{t}$,
    ]

    \nextgroupplot[ylabel=density $\rho$]
    \addplot[NavyBlue, dashed, opacity=0.9] coordinates {
    (-5,1.9) (0,1.9) (0,0.2) (5,0.2)
    };
    \addplot[NavyBlue, very thick] table [x=s,y=density, col sep=comma] {data/iso_riemann_solution.csv};
    \nextgroupplot[ylabel=momentum $m$]
    \addplot[NavyBlue, dashed, opacity=0.9] coordinates {
    (-5,0) (0,0) (0,0) (5,0)
    };
    \addplot[NavyBlue, very thick] table [x=s,y=momentum, col sep=comma] {data/iso_riemann_solution.csv};
\end{groupplot}
\end{tikzpicture}%
}
\caption{Reference Riemann solution for the isothermal two-phase flow model problem.
The dashed lines represent the initial conditions.}
\label{fig:riemann_solution:iso}
\end{figure}
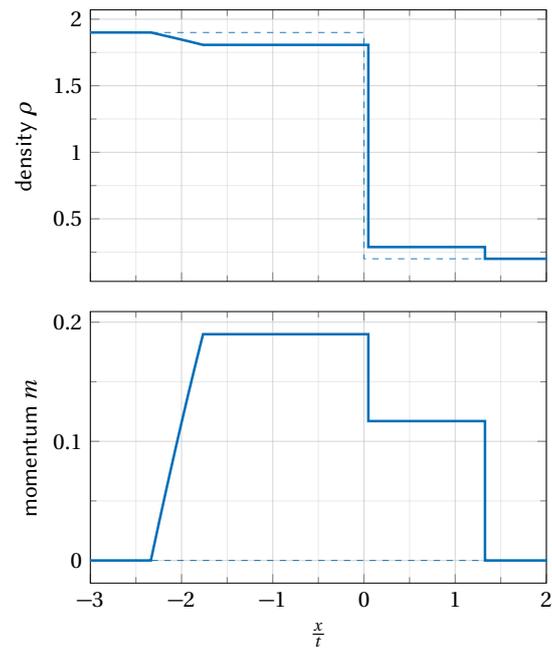

\begin{figure}[H]
\centering
\makebox[\textwidth][c]{
\includegraphics[width=1.05\columnwidth]{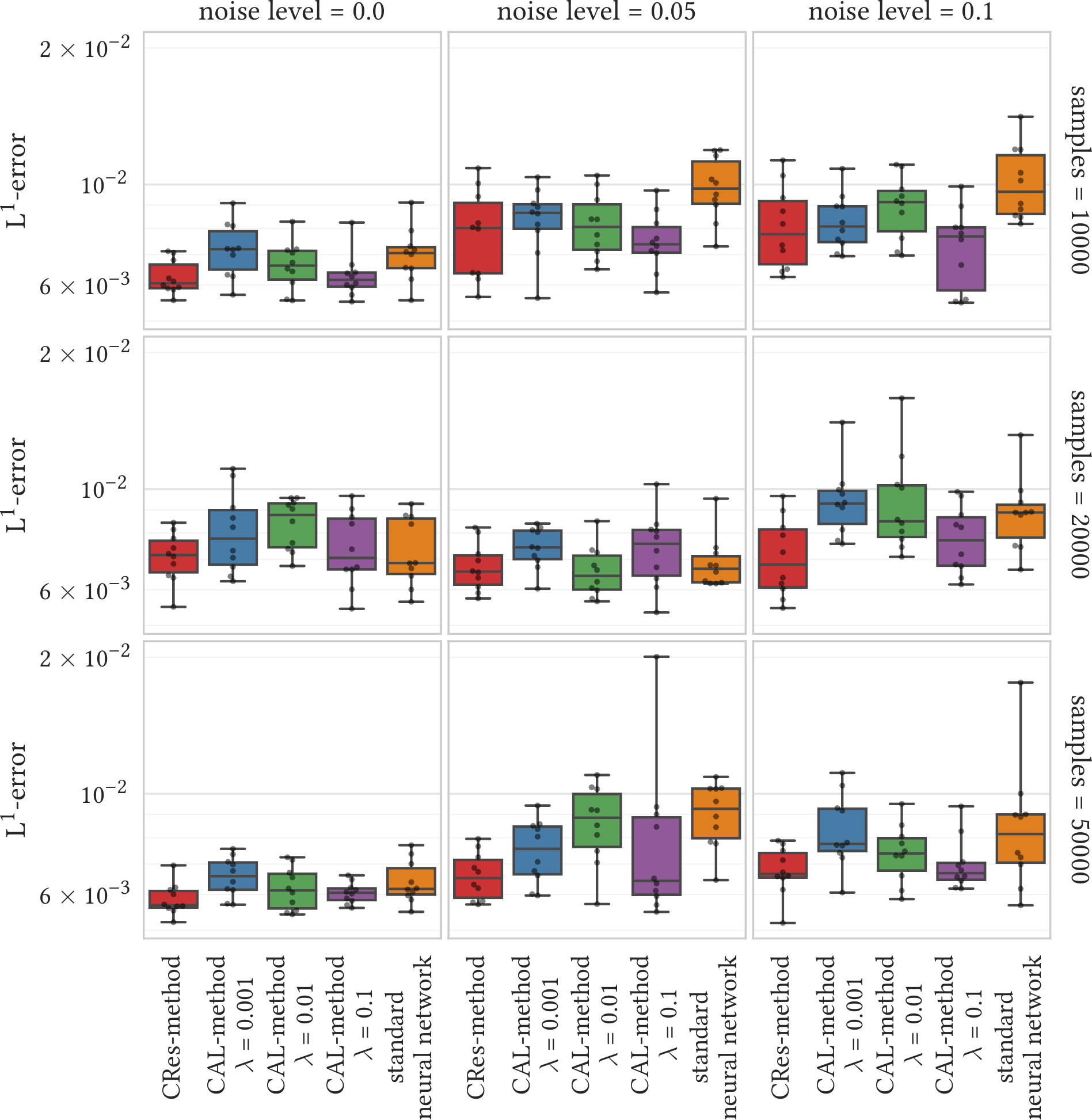}
}
\caption{Mean $L^1$-errors for the two-phase flow model problem.
The mean $L^1$-errors of the finite volume simulation are computed over \num{25} perturbed Riemann problems for each trained network.
The gray dots represent the error for each network instance.
}
\label{fig:continuum_error:iso}
\end{figure}

As the second test, we apply the networks in simulations.
We consider the Riemann problem with $\rho_\phasel = 1.9$, $\rho_\phaser = 0.2$, and $m_\phaselr = 0$.
The corresponding Riemann solution is depicted in \autoref{fig:riemann_solution:iso}.
The simulations are performed on a grid with \num{2000} uniform cells on the domain $[-1, 1]$.
The timestep is $\Delta t = \num{e-4}$, the final time is $t_{\mathrm{end}} = \num{0.25}$ and the Lax--Friedrichs parameter is $\alpha_{\mathrm{LF}} = \num{2}$.
The numerical error of the discretization, i.e. if the exact solver is used in \eqref{eq:interface_flux:right} and \eqref{eq:interface_flux:left}, is approximately \num{5.2e-3}.
\newline
Again, we average the error over \num{25} simulations.
For that we perturb the initial densities $\rho_\phaselr$ randomly by 2\% of their value, i.e.
we set $\rho_\phaselr \cdot (1 + \num{0.02} \cdot R)$ as the initial values, where $R$ is an uniformly distributed random variable on $[-1,1]$.
\newline
The $L^1$-error of the simulation with respect to the analytic solution is shown in \autoref{fig:continuum_error:iso} for the different networks.
We see that the $L^1$-error in case of noise-free data is similar across all methods.
However for higher noise levels, the results are more varied.
For all, except one data set, the \cres-method gives lower errors on average than the standard networks.
The performance of the \cadl-method depends highly on the choice of the constraint penalty parameter $\lambda$; depending on this choice, the \cadl-method can perform better than the standard network on noisy data sets.
\newline
Between the \cres- and the \cadl-method, the former gives more reliable results, and it does not require selecting a parameter.

\pagebreak
\subsection{Euler Equations}
\label{sec:euler_results}

Just as in the previous two sections, we investigate the constraint-aware training methods --- this time for the Euler model.
The data set consists of random values $\rho_{\phaselr} \in [\num{0.05}, \num{1.3}]$, $m_{\phaselr} \in [\num{-1.95}, \num{1.95}]$, and $E_{\phaselr} \in [\num{0.11875}, \num{7.9625}]$, which have been sampled as described in \autoref{sec:data_generation}.
All data points are discarded that do not fulfill the positivity condition, i.e., Riemann data that produce a vacuum, see \cite{toro:riemann:2009}.
The resulting data sets have $N_{\mathrm{train}} \in \{\num{10000},~\num{20000}~\num{50000}\}$ samples, with noise levels $l_{\mathrm{noise}} \in \{\num{0},~\num{0.05}\}$.
The test data $D_{\mathrm{test}}$ consists of \num{10986} samples --- after discarding invalid samples according to \eqref{eq:positivity_condition} --- on a uniform grid over $B$.

\begin{figure}[H]
\centering
\includegraphics[width=0.8\columnwidth]{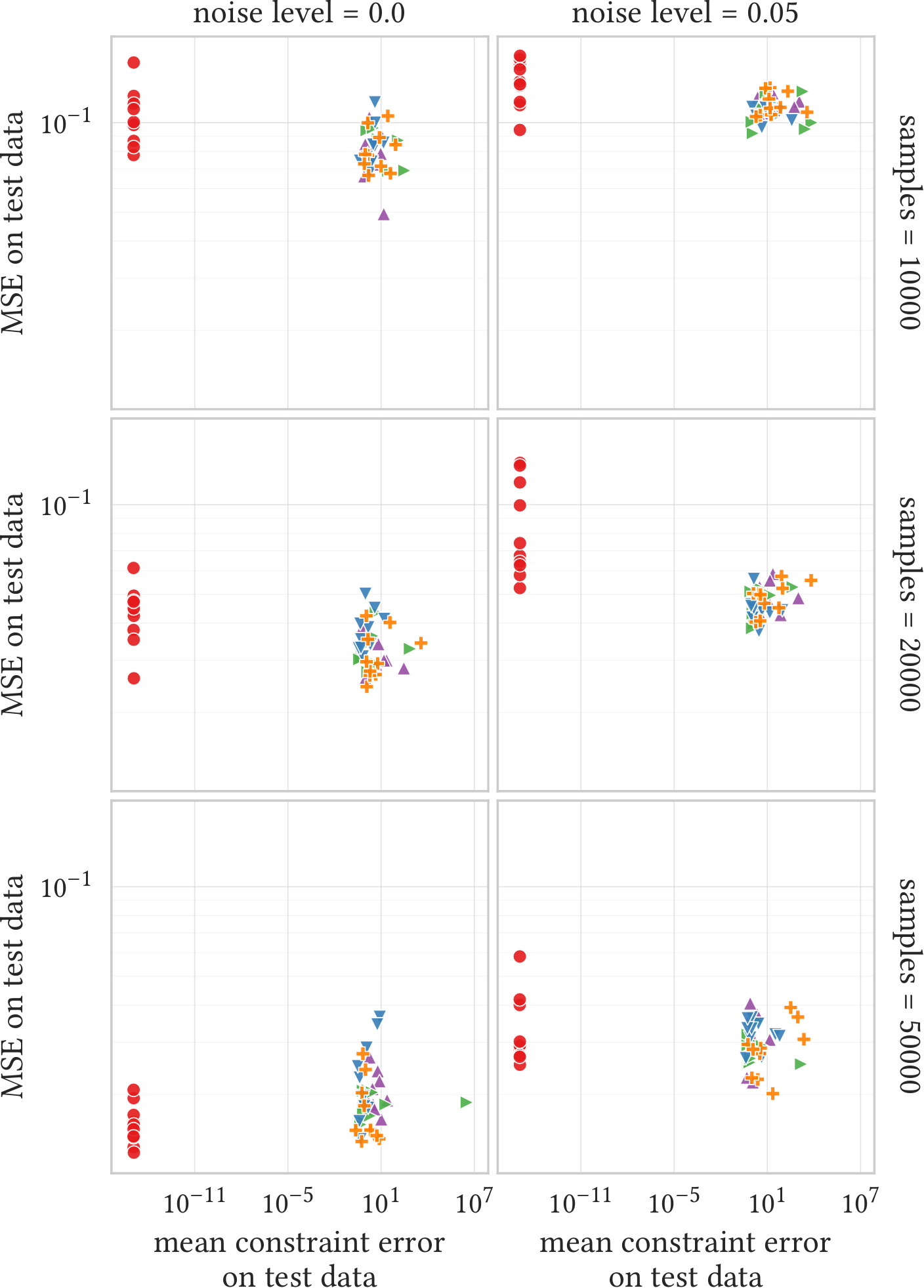}
\mbox{}\\[2ex]
\includegraphics[width=0.9\columnwidth]{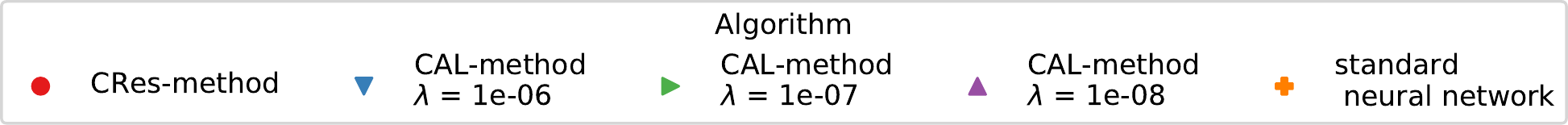}
\caption{Mean-squared errors and constraint errors on the test data set for the Euler model problem.
Each subfigure corresponds to one data set. \\
Horizontally the mean squared error on the test data set is plotted, vertically the mean constraint error.
Therefore, points that are located to the left and near bottom are considered better.
}
\label{fig:nn_benchmark:euler}
\end{figure}

The results are depicted in \autoref{fig:nn_benchmark:euler}.
Again, the \cres-method clearly outperforms the standard networks and the \cadl-method regarding the constraint error.
Concerning the standard networks and the \cadl-method no difference is discernible.
This is due to the fact that a much smaller constraint penalty parameter is used than for the previous two model problems.
Tests with higher values of $\lambda$ did not show a significant decline in the training error, from which we can concur that the optimization problem is seemingly much harder compared to the other two model problems.

\begin{figure}[H]
 \centering
\resizebox{0.5\columnwidth}{!}{
\tikzset{external/export next=false}
\tikzsetnextfilename{riemannsolution_euler}
\begin{tikzpicture}[scale=1.0, font=\sffamily]
\begin{groupplot}[
    group style={
        group name=my plots,
        group size=1 by 3,
        x descriptions at=edge bottom,
        y descriptions at=edge left,
        horizontal sep=0.5cm,
        vertical sep=0.5cm,
    },
    width=0.5\columnwidth, height=0.3\columnwidth,
    scale only axis,
    grid=both, minor x tick num=1, minor y tick num=1,
    major grid style={very thin, opacity=0.66},
    minor grid style={ultra thin, opacity=0.33},
    xmin=-2, xmax=2.5,
    xlabel=$\frac{\spx}{t}$,
    ]

    \nextgroupplot[ylabel=density $\rho$]
    \addplot[NavyBlue, dashed, opacity=0.9] coordinates {
    (-5,1.0) (0,1.0) (0,0.125) (5,0.125)
    };
    \addplot[NavyBlue, very thick] table [x=s,y=density, col sep=comma] {data/euler_riemann_solution.csv};

    \nextgroupplot[ylabel=momentum $m$]
    \addplot[NavyBlue, dashed, opacity=0.9] coordinates {
    (-5,0) (0,0) (0,0) (5,0)
    };
    \addplot[NavyBlue, very thick] table [x=s,y=momentum, col sep=comma] {data/euler_riemann_solution.csv};

    \nextgroupplot[ylabel=total energy $E$]
    \addplot[NavyBlue, dashed, opacity=0.9] coordinates {
    (-5, 2.5) (0, 2.5) (0,0.25) (5,0.25)
    };
    \addplot[NavyBlue, very thick] table [x=s,y=total energy, col sep=comma] {data/euler_riemann_solution.csv};
\end{groupplot}
\end{tikzpicture}%
}
\caption{Reference Riemann solution for the Euler model problem. The dashed lines represent the initial conditions.}
\label{fig:riemann_solution:euler}
\end{figure}
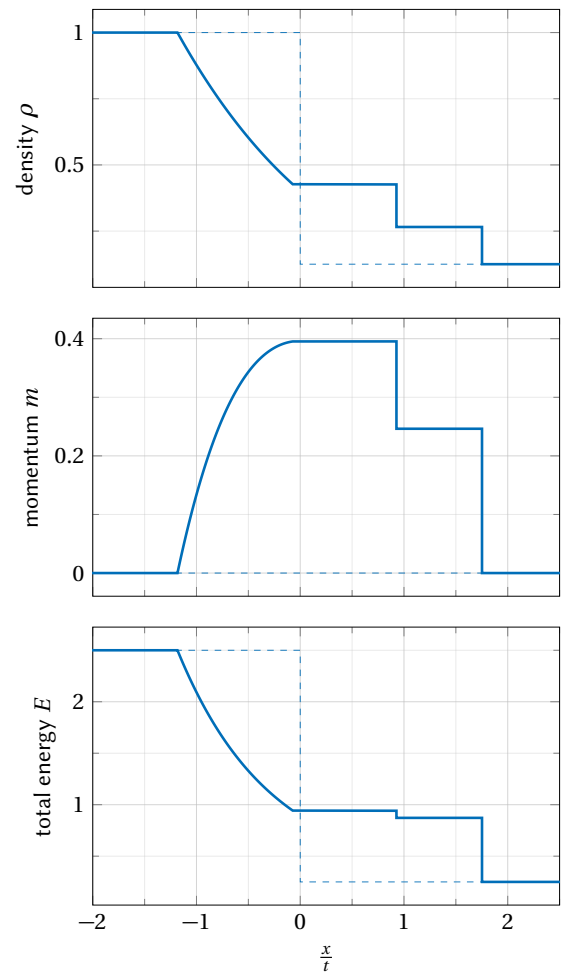

\begin{figure}[H]
\centering
\includegraphics[width=0.8\columnwidth]{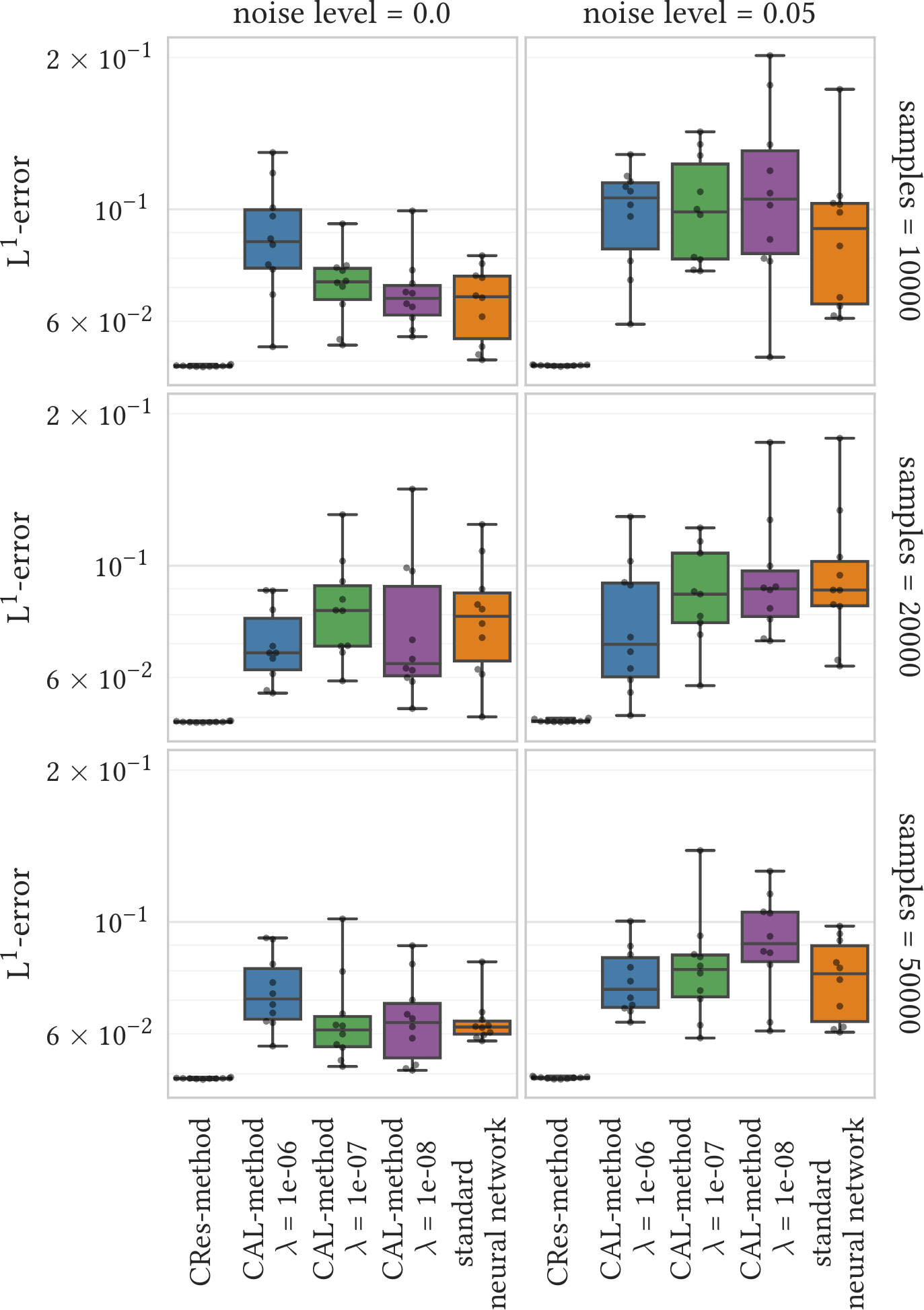}
\caption{Mean $L^1$-errors for the Euler model problem.
The mean $L^1$-errors of the finite volume simulation are computed over \num{25} perturbed Riemann problems for each trained network.
The gray dots represent the error for each network instance.
}
\label{fig:continuum_error:euler}
\end{figure}

For the continuum simulation tests, we consider the initial conditions
$\rho_\phasel = \num{1}$, $\rho_\phaser = \num{0.125}$, $m_\phaselr = 0$, $E_\phasel = \num{2.5}$, and $E_\phaser = \num{0.25}$.
which correspond to the Sod shock tube benchmark example \cite{sod:survey:1978}.
The analytic solution is shown in \autoref{fig:riemann_solution:euler}
\newline
The mesh consists of \num{1000} cells on the domain $(-1, 1)$.
The timestep is $\Delta t = \num{5e-5}$, the final time is $t_{\mathrm{end}} = \num{0.25}$ and the Lax--Friedrichs parameter is $\alpha_{\mathrm{LF}} = \num{6}$.
\newline
If we apply the exact Riemann solver for the numerical interface fluxes \eqref{eq:interface_flux:right}, \eqref{eq:interface_flux:left}, the  $L^1$-discretization error is approximately \num{0.09}.
\newline
As before, we perturb the initial condition densities by 1\% and measure the $L^1$-errors w.r.t. the corresponding exact Riemann solutions and average the $L^1$-error over \num{25} simulations for each method and data set.
\newline
The averaged $L^1$-errors are depicted in \autoref{fig:continuum_error:euler}.
We can see that the \cres-method continues to work best.
The \cadl-method on the other hand is often even worse than the standard approach, despite choosing a relatively small constraint penalty parameter $\lambda$.
This is due to the intrinsic high dimensional input space of the constraint \eqref{eq:constraint:euler}, which consequently makes the training with the loss function \eqref{eq:constraint_adapted_loss} much harder.

\section{Discussion}
\label{sec:summary}

In this work we investigated two different methods (\cres{} and \cadl) to build constraint-aware neural networks and their application to numerical simulations.
The performance of the methods was examined on the basis of three different model problems: the scalar cubic flux model problem (\autoref{sec:cubic}), the isothermal two-phase flow model problem (\autoref{sec:iso_euler}) and the Euler model problem (\autoref{sec:euler_model}).

From the results in \autoref{sec:results} we can draw several conclusions.
First, we note that Riemann solvers based on neural networks with a low constraint error yield a high numerical accuracy in numerical simulations.
\newline
Secondly, the \cres-method is able to fulfill a constraint perfectly if an exact resolving function $\Psi$, such as \eqref{eq:psi:cubic} or \eqref{eq:psi:euler}, is used --- see \autoref{sec:cubic_results} and \autoref{sec:euler_results}.
Nonetheless, even if $\Psi$ is just an approximation,  like \eqref{eq:psi:iso}, the \cres-method outperforms the standard approach and the \cadl-method with respect to the constraint error, as shown in \autoref{sec:iso_results}.
This however comes at the cost that the constraint has to be handled analytically.
We stress that the \cres-method is more than merely a post-processing step, as the full information from the training set can be used and the knowledge about the constraint is incorporated in the training of the network.
\newline
On the other hand, the \cadl-method is much more generic.
It requires only a constraint function \eqref{eq:constraint:rh} and includes the deviation from the constraint as a penalty term in the training procedure.
For the cubic flux model problem and the isothermal two-phase flow model problem this method works well, in particular for high noise levels.
In case of the Euler model problem with the constraint \eqref{eq:constraint:euler}, the \cadl-method seems to reach it limits, due to the high dimensionality of the constraint.
Standard optimization algorithms such as stochastic gradient descent are apparently not powerful enough to resolve a complex constraint such as \eqref{eq:constraint:euler} when they are included in the loss function \eqref{eq:constraint_adapted_loss}.

There remain open questions and possible improvements for the proposed methods.
Concerning the \cadl-method, one could apply dynamic scheduling for the constraint penalty parameter $\lambda$, e.g. starting with $\lambda = 0$ and slowly increasing the parameter in each epoch.
Another option would be to train the network without considering any constraints, and then retrain the model with a constraint-aware method.
\newline
Furthermore, choosing the square norm --- or other nonlinear scalings --- of the constraint in \eqref{eq:constraint_loss} might result in a different performance of the resulting networks and should be investigated.

\subsection*{Acknowledgements}
The work was supported by the German Research Foundation (DFG) through SFB TRR 75 ``Droplet dynamics under extreme ambient conditions''.
\printbibliography

\appendix
\section{Riemann Solver for the Cubic Flux Model with Kinetic Relation}
\label{sec:cubic_riemannsolver}

For the cubic flux model \eqref{eq:cubic_cl} we want to solve the Riemann problem
\begin{align}
 u(x,0) = \begin{cases}
           u_{\phasel} & \text{ for } x < 0, \\
           u_{\phaser} & \text{ for } x > 0,
          \end{cases}
\end{align}
with $u_{\phasel} > 0$, while selecting solutions that satisfy the kinetic relation \eqref{eq:kinrel}.
The kinetic relation comes with a corresponding, so-called companion function (see \cite{lefloch:hyperbolic:2002}), which is defined in this instance by
\begin{align} \label{eq:companion_function}
 \varphi^\sharp(u) = - (1 - \kappa) u.
\end{align}
For the scalar problem \eqref{eq:cubic_cl} it is possible to compute the wave speed
\begin{align} \label{eq:scalar_wave_speed:appendix}
 s = \overline{s}(u_{\phaser}, u_{\phasel}) \coloneqq
 \begin{cases}
 \dfrac{f(u_{\phaser}) - f(u_{\phasel})}{u_{\phaser} - u_{\phasel}} & \text{ for } u_{\phaser} \neq u_{\phasel}, \\
 f'(u_{\phaser}) & \text{ if } u_{\phaser} = u_{\phasel}.
 \end{cases}
\end{align}
The Riemann solver for this problem was presented in \cite{chalons.engel.ea:conservative:2014} and distinguishes between five different cases:
\begin{enumerate}[label=\protect\circled{\arabic*}]
 \item \label{item:rs:1} For $u_{\phaser} \geq u_{\phasel}$ the Riemann solution is a classical rarefaction wave
 \begin{align*}
   u(x,t) = \begin{cases}
           u_{\phasel}
           & \text{ for }  \tfrac{x}{t} \leq f'(u_{\phasel}), \\
           \Bigl(f'\big\vert_{[f'(u_{\phasel}), f'(u_{\phaser})]} \Bigr)^{-1} (\tfrac{x}{t})
           & \text{ for }  f'(u_{\phasel}) \leq \tfrac{x}{t} \leq f'(u_{\phaser}), \\
           u_{\phaser}
           & \text{ for }  \tfrac{x}{t} > f'(u_{\phaser}).
          \end{cases}
 \end{align*}
 \item \label{item:rs:2} For $u_{\phaser} \in [0, u_{\phasel})$ the solution is a single Laxian shock wave
 \begin{align*}
 u(x,t) = \begin{cases}
           u_{\phasel}
           & \text{ for }  \tfrac{x}{t} \leq \overline{s}(u_{\phasel}, u_{\phaser}), \\
           u_{\phaser}
           & \text{ for } \tfrac{x}{t} > \overline{s}(u_{\phasel},u_{\phaser}).
          \end{cases}
\end{align*}
\item \label{item:rs:3} For $u_{\phaser} \in [\varphi^\sharp(u_{\phasel}), 0)$ the interface is also a single Laxian shock wave
 \begin{align*}
 u(x,t) = \begin{cases}
           u_{\phasel}
           & \text{ for }  \tfrac{x}{t} \leq \overline{s}(u_{\phasel}, u_{\phaser}), \\
           u_{\phaser}
           & \text{ for } \tfrac{x}{t} > \overline{s}(u_{\phasel},u_{\phaser}).
          \end{cases}
\end{align*}
\item \label{item:rs:4} For $u_{\phaser} \in \varphi(u_{\phasel}, \varphi^\sharp(u_{\phasel})$:
If $\varphi(u_{\phasel}) \neq \varphi^\sharp(u_{\phasel}) \neq \varphi^{\mathrm{char}}(u_{\phasel})$ with $\varphi^{\mathrm{char}}(u) \coloneqq -\tfrac{1}{2} u$, the solution consists of a Laxian shock wave followed by an undercompressive interface:
 \begin{align*}
 u(x,t) = \begin{cases}
           u_{\phasel}
           & \text{ for }  \tfrac{x}{t} < \overline{s}(u_{\phasel}, \varphi(u_{\phasel}) ), \\
           \varphi(u_{\phasel})
           & \text{ for }  \overline{s}(u_{\phasel}, \varphi(u_{\phasel}) ) \leq \tfrac{x}{t} \leq \overline{s}(\varphi(u_{\phasel}), u_{\phaser} ), \\
           u_{\phaser}
           & \text{ for } \tfrac{x}{t} > \overline{s}(\varphi(u_{\phasel}),u_{\phaser}).
          \end{cases}
\end{align*}
If $\varphi(u_{\phasel}) = \varphi^\sharp(u_{\phasel}) = \varphi^{\mathrm{char}}(u_{\phasel})$ the solution is a single characteristic shock wave from $u_{\phasel}$ to $u_{\phaser}$.

\item \label{item:rs:5} For $u_{\phaser} \in (- \infty, \varphi(u_{\phasel}) )$:
If $\varphi(u_{\phasel}) \neq \varphi^\sharp(u_{\phasel}) \neq \varphi^{\mathrm{char}}(u_{\phasel})$, the undercompressive interface follows a rarefaction wave, i.e. the solution is given by
\begin{align*}
  u(x,t) = \begin{cases}
           u_{\phasel}
           & \text{ for }  \tfrac{x}{t} < \overline{s}(u_{\phasel}, \varphi(u_{\phasel}) ), \\
           \varphi(u_{\phasel})
           & \text{ for }   \overline{s}(u_{\phasel}, \varphi(u_{\phasel}) )
           \leq \tfrac{x}{t}
           \leq f'(\varphi(u_{\phasel})), \\
           \Bigl(f'\big\vert_{\bR^-} \Bigr)^{-1} (\tfrac{x}{t})
           & \text{ for }   f'(\varphi(u_{\phasel}))
           \leq \tfrac{x}{t}
           \leq  f'(u_{\phaser}), \\
            u_{\phaser} & \text{ for }  \tfrac{x}{t} >  f'(u_{\phaser}).
          \end{cases}
\end{align*}
If $\varphi(u_{\phasel}) = \varphi^\sharp(u_{\phasel}) = \varphi^{\mathrm{char}}(u_{\phasel})$ the solution consists of a characteristic shock wave from $u_{\phasel}$ to $\varphi^{\mathrm{char}}(u_{\phasel})$ attached to a rarefaction wave to $u_{\phaser}$.
\end{enumerate}

\end{document}